\documentclass[a4paper,11pt]{article}
\pdfoutput=1 % if your are submitting a pdflatex (i.e. if you have
\usepackage{jcappub} % for details on the use of the package, please
\usepackage[T1]{fontenc} % if needed
\usepackage{mathtools}

\subheader{\footnotesize
        \vspace*{-3.7em}
        \begin{flushright}
                CERN-TH-2019-222 \\
                MITP/19-087 \\
                INR-TH-2019-025
        \end{flushright}
}

\usepackage{aas_macros}
\usepackage{amsmath,amssymb}
\usepackage{cleveref}
\usepackage{siunitx}
\usepackage{enumerate}
\usepackage[dvipsnames]{xcolor}

\definecolor{darkgreen}{rgb}{0,0.5,0}

\newcommand{\beq}{\begin{eqnarray}}
\newcommand{\eeq}{\end{eqnarray}}

\newcommand{\ev}[1]{\ensuremath{\left\langle #1 %
                     \right\rangle}} % Expectation value
\title{Looking for MACHOs in the Spectra of Fast Radio Bursts}
\author[a,b]{Andrey~Katz,}
\author[a,c]{Joachim~Kopp,}
\author[a,d,e]{Sergey~Sibiryakov,}
\author[f,a]{and Wei~Xue}

\affiliation[a]{Theoretical Physics 
Department, CERN, \\
1 Esplanade des Particules, 
CH-1211 Geneva 23, Switzerland} 
\affiliation[b]{D\'epartement de Physique Th\'eorique and
                Center for Astroparticle Physics (CAP),\\
                Universit\'e de Gen\`eve, 24 quai Ansermet, CH-1211
                Gen\`eve 4, Switzerland} 
\affiliation[c]{PRISMA Cluster of Excellence \& Mainz Institute for
  Theoretical Physics, \\ 
                Johannes Gutenberg University, Staudingerweg 7, 55099
                Mainz, Germany} 
\affiliation[d]{Institute of Physics, Laboratory for Particle Physics
  and Cosmology (LPPC), \\
\' Ecole Polytechnique F\' ed\' erale de Lausanne (EPFL), CH-1015
Lausanne, Switzerland}
\affiliation[e]{Institute for Nuclear Research of the Russian Academy
  of Sciences, \\ 
                60th October Anniversary Prospect, 7a, 117312 Moscow, Russia}
\affiliation[f]{Department of Physics, University of Florida,
  Gainesville, FL 32611, USA} 

\emailAdd{andrey.katz@cern.ch}
\emailAdd{joachim.kopp@cern.ch}
\emailAdd{sergey.sibiryakov@cern.ch}
\emailAdd{weixue@ufl.edu}

\abstract{We explore a novel search strategy for dark matter in the form of
  massive compact halo objects (MACHOs) such as primordial black holes or dense
  mini-halos in the mass range from \SI{e-4}{M_\odot} to~\SI{0.1}{M_\odot}.
  These objects can gravitationally lens the signal of fast radio bursts
  (FRBs), producing a characteristic interference pattern in the frequency
  spectrum, similar to the previously studied femtolensing signal in gamma ray
  burst spectra.  Unlike traditional searches using microlensing, FRB lensing
  will probe the abundance of MACHOs at cosmological distance scales ($\sim
  \si{Gpc}$) rather than just their distribution in the neighborhood of the
  Milky Way.  The method is thus particularly relevant for dark mini-halos,
  which may be inaccessible to microlensing due to their finite spatial extent
  or tidal disruption in galaxies.  We find that the main complication in FRB
  lensing will be interstellar scintillation in the FRB's host galaxy and in
  the Milky Way. Scintillation is difficult to quantify because it heavily
  depends on turbulence in the interstellar medium, which is poorly understood.
  We show that, nevertheless, for realistic scintillation parameters, FRB
  lensing can set competitive limits on compact dark matter object, and we back
  our findings with explicit simulations.
}

\begin{document}
\maketitle
\flushbottom

%==========================================================================
\section{Introduction}
\label{sec:intro}
%==========================================================================

%  !TEX root =  draft.tex
Massive compact halo objects (MACHOs) have been widely discussed in the
literature as possible dark matter candidates. In particular, a lot of
attention has been paid to two classes of objects: primordial black holes
(PBHs) \cite{Hawking:1971ei, Carr:2016drx, Sasaki:2018dmp} and ultra-compact mini-halos (UCMHs)
\cite{Hogan:1988mp, Kolb:1993zz, Kolb:1993hw, Zurek:2006sy, Hardy:2016mns}.\footnote{For
recent N-body simulations of UCMHs in the context of dark matter composed of
the QCD axion see~\cite{Vaquero:2018tib, Buschmann:2019icd, Eggemeier:2019khm}.} For
concreteness, we will focus our discussion on the case of PBHs, though most
of our findings will apply also to other types of MACHOs.

The interesting mass range for PBHs is between
\SI{e-16}{M_\odot} and \SI{e3}{M_\odot} (\SI{e17}{grams} to
\SI{e36}{grams}), with ${\rm M}_\odot$ being the solar mass.  On the lower end
of this range, the PBH abundance is strongly constrained by non-observation of
the $\gamma$-ray flux that would be produced by PBHs due to their Hawking
evaporation~\cite{Carr:2009jm}, and by measurements of the cosmic microwave
background (CMB) anisotropies that would be affected by Hawking radiation in
the early universe~\cite{Poulter:2019ooo}.  At large PBH masses, constraints
are due to accretion onto PBHs, which would again affect the CMB
anisotropies~\cite{Ali-Haimoud:2016mbv}.  A massive effort has been put into
exploring the available parameter space of PBH dark matter, in particular via
microlensing searches like MACHO~\cite{Allsman:2000kg},
EROS~\cite{Tisserand:2006zx}, OGLE~\cite{Wyrzykowski:2011tr}, and most recently
SUBARU-HSC~\cite{Niikura:2017zjd}. Other methods to search for MACHOs in the
abovementioned mass range exist, but their sensitivity is typically inferior to
microlensing. Note also that several published constraints have not withstood
more careful analysis and are therefore now considered invalid or
controversial.  Thus, big portions of MACHO parameter space remain
unconstrained, while others are only loosely constrained and can still
accommodate a sizable fraction of dark matter in the form of MACHOs.

It is important to note that existing microlensing searches --- which are responsible
for essentially all the non-controversial constraints on MACHOs in the
interesting mass range --- are based on observations in the Milky Way or the
Local Group of galaxies. Drawing conclusions on the MACHO abundance on
cosmological scales thus requires extrapolation, which suffers from several
uncertainties. First, translating observational results into constraints on the
MACHO abundance requires assumptions about the dark matter density
profile of the Milky Way, and for some observations like those by SUBARU-HSC,
of the other galaxies in the Local Group. Second, the mass function of MACHOs
and their abundance in the Local Group may differ from those on cosmological
scales. Indeed, for the case of PBHs, one can expect the mass distribution to
evolve in time due to accretion and mergers. Composite
MACHOs such as UCMHs can be tidally disrupted inside big galactic
halos~\cite{Zhao:2005py, Berezinsky:2005py, Green:2006hh, Goerdt:2006hp,
Berezinsky:2014wya, Tinyakov:2015cgg, Fairbairn:2017sil, Dokuchaev:2017psd}
so that more of them can be found at
larger redshifts and in less dense objects than in today's Milky Way.
Finally, microlensing searches cannot constrain UCMHs or other MACHOs
of finite extent if their size exceeds their Einstein radius at galactic
distances \cite{Dror:2019twh}.

In this paper, we investigate a search strategy that probes MACHOs directly at
cosmological distance scales. It is based on an effect which we will call {\em
diffractive gravitational lensing}. The basic idea of diffractive lensing is
that a MACHO placed between a distant source and the observer and whose mass is
too small to produce two resolved images of the source still causes a phase
difference between the unresolved images. This leads to characteristic
interference patterns in the energy spectrum of the source.  Diffractive
lensing was originally proposed in the context of gamma-ray burst (GRB)
sources, where it is known as femtolensing~\cite{Gould:1991td,
1993ApJ...413L...7S}.\footnote{The term ``femtolensing'' refers to the typical
angular separation of order a femto-arcsecond between the two images produced
by the lenses considered in the context of GRBs. In this paper we consider much
heavier lenses corresponding to larger angular separations between images, so
we prefer to designate the effect by the generic term ``diffractive lensing''.}
Diffractive lensing of GRBs could be expected to constrain MACHOs in the mass
range from \num{e-17} to \SI{e-14}{M_\odot}~\cite{Barnacka:2012bm}.  However,
due to the non-negligible angular sizes of typical GRBs, it turns out that the
interference pattern in the spectrum would be washed
out~\cite{Matsunaga:2006uc}, so that femtolensing does not currently yield
bounds on the MACHO abundance~\cite{Katz:2018zrn}.

We propose here to apply the diffractive lensing technique to fast radio bursts
(FRBs, see e.g. reviews \cite{Katz:2018xiu, Popov:2018hkz, Petroff:2019tty}) rather than GRBs.
As the photon wavelengths for FRBs are much longer than for GRBs, the method
will be sensitive to much heavier MACHOs, namely those between \num{e-4} and
\SI{0.1}{M_\odot}.  The Einstein radii of MACHOs in this mass range are much
larger than those of the lenses considered in the context of GRBs, therefore
the requirement on the source size is much less stringent and easily satisfied
by putative FRB progenitors.

Gravitational lensing of FRBs has been considered previously in the literature:
strong lensing, in particular the observation of two consecutive bursts
separated in time, has been proposed in~\cite{Munoz:2016tmg} and further
considered in~\cite{Dai:2017twh, Laha:2018zav}. Such an observation would be
sensitive to MACHOs of \SI{10}{M_\odot} to \SI{100}{M_\odot}.  Diffractive
lensing of FRBs has been suggested in~\cite{Zheng:2014rpa, Eichler:2017eid},
where it was also pointed out that the main complication in applying this
technique is the distortion of FRB spectra by
scintillation~\cite{Rickett:1977vv, doi:10.1146/annurev.aa.28.090190.003021,
1992RSPTA.341..151N}.  The latter appears due to scattering of radio waves
on the turbulent interstellar medium (ISM), leading to multi-path propagation
of radio signals and thus to random interference patterns in the frequency
spectrum. Nevertheless, ref.~\cite{Eichler:2017eid} argued that the lensing
signal can be disentangled from scintillation by considering the temporal
autocorrelation of the electromagnetic field of the radio wave.

In the present paper we take further steps to explore this idea.  We show that
one does not need to measure the full electromagnetic field amplitude to
extract the lensing signal or confirm the absence thereof. In fact, the entire
analysis can be carried out in terms of the intensity (fluence) spectrum of the wave,
without any need for phase information. This has practical importance as
most FRB detections occur in survey mode where the full field information may
not be recorded. We also demonstrate that the extraction of the lensing signal
from scintillation is possible whenever these two phenomena are characterized
by disparate frequency scales. Apart from the regime considered
in~\cite{Eichler:2017eid} when the two lensed images are unresolved by the
scintillation screen, this also includes the case when the scintillation screen
does resolve the images and distorts them incoherently, provided the lensing
time delay is longer than the inverse of the decorrelation bandwidth of the
scintillation.  Finally, we take into account scintillation in the
intergalactic medium (IGM). For realistic parameter choices we find IGM
scintillation to be less relevant than scintillation in the Milky Way and in
the host galaxy of the source, as long as the line-of-sight does not cross any
major concentrations of ionized plasma, such as galaxy clusters. Our analytic
estimates are supported by numerical simulations of FRB spectra with lensing
and scintillation, using a simplified model in which the ISM and IGM
are described by one-dimensional scintillation screens.

We propose a data analysis procedure to search for the lensing signal by
looking for peaks in the Fourier transform of the FRB intensity spectrum. We
find that the sensitivity is improved if the smooth component of the spectrum
is first divided out in order to reconstruct an approximation to the transfer
function (the ratio between the emitted and recorded intensity).  This
procedure also reduces our sensitivity to imperfect
modeling of the FRB spectrum.  The search for the peaks in the Fourier
transform of the transfer function is similar to the one used for resonance
searches on a smooth background in collider experiments. We validate our
procedure using the simulated data.
 
A precise modeling of scintillation is problematic due to the poor
understanding of the turbulent ISM and IGM, leading to large uncertainties in
the parameters governing scintillation. In spite of these complications, we
will see below that there is realistic hope of extracting the lensing signal
from FRB data.  Moreover, the reach of the method will significantly improve in
the future with an increased number of FRB detections, in particular thanks to
the CHIME and SKA telescopes, which are expected to detect a few tens of FRBs
per day~\cite{Chawla:2017wdz, Macquart:2015uea}.  For the time being, we will
explore the sensitivity as a function of the scintillation parameters and the
number of FRBs.

The paper is organized as follows. In~\cref{sec:femto} we briefly review the
formalism of gravitational diffractive lensing as well as the potential and the
limitations of MACHO searches using this technique. In~\cref{sec:scintillation}
we address the problem of scintillation. We review the relevant properties of
the ISM and IGM, and we analytically estimate the impact of scintillation on
diffractive lensing of FRBs.  We then proceed in \cref{sec:simulation-and-results}
to the discussion of numerical results obtained by explicitly simulating the
propagation of FRB signals.  In this section, we also present our main results
in the form of sensitivity estimates.  We conclude in \cref{sec:conclusions}.

%==========================================================================
\section{Review of Diffractive Lensing}
\label{sec:femto}
%==========================================================================

%  !TEX root =  draft.tex
Consider a distant source such as an FRB, and a gravitational lens close to the
line of sight.  For our purposes, it is sufficient to consider the source and the
lens to be point-like. 
In general, a point-like lens produces two images
of the source, but if the lens has a small mass, the observer will not be able to resolve
them.  However, the photons corresponding to the two images travel different
distances and experience different gravitational potentials, so their travel
times will differ by an amount $\Delta t$. This leads to a relative phase shift
$\Delta \phi = \omega \, \Delta t $, where $\omega$ is the photon energy.  If
$\Delta\phi \gtrsim \mathcal{O}(1)$, an interference pattern  
will appear in the photon energy
spectrum, and may allow detection of the lens even when the two images are
not resolved.

More quantitatively, the time delay between the images is given
by\footnote{We work in the units $c=\hbar=1$ and assume that the background
spactime is described by the spatially flat Friedmann--Robertson--Walker metric.}
\begin{align}
  \Delta t = \frac{D_L D_S}{D_{LS}}
             \bigg( \frac{|\vec \theta - \vec \beta|^2}{2} - \psi(\vec
             \theta) \bigg)\;, 
  \label{eq:deltatgeom}
\end{align}
where $D_L$, $D_S$, $D_{LS}$ are the \emph{comoving} distances
between the observer and the lens, between the observer and the source,
and between the lens and the source, respectively;
$\beta$ is the angle under which the observer would
see the source in the absence of the lens, whereas $\theta$ is the
angle of the lensed image; 
$\psi(\vec\theta)$ stands for the lensing potential, which depends on the
density profile of the lens. The lensing potential of a point-like mass $M$
is
\begin{align}
  \psi(\vec\theta) = \theta_E^2 \log|\vec\theta| \,,
  \label{eq:psi-point}
\end{align}
where the Einstein angle
\begin{align}
  \theta_E \equiv \bigg( 4 GM (1+Z_L) \frac{D_{LS}}{D_L D_S} \bigg)^{1/2}
  \label{eq:EinsteinAngle}
\end{align}
is a measure for the typical angle under which the images are observed
relative to the lens.  In \cref{eq:EinsteinAngle}, $Z_L$ denotes the
redshift of the lens.  The positions of
the images are given by the solutions to the lens equation,
\begin{align}
  \vec\theta - \vec\beta = \vec\nabla\psi(\vec\theta) \,,
  \label{eq:lens-equation}
\end{align}
which for a point-like lens gives two locations
\begin{align}
  \theta_\pm = \frac{1}{2} \Big( \beta \pm \sqrt{\beta^2 + 4
    \theta_E^2} \Big) \,. 
  \label{eq:theta-pm}
\end{align}
Different signs of the two angles here mean that the images occur on
opposite sides of the lens.
Denoting $y \equiv \beta / \theta_E$, the magnifications of the
two images are \cite{Bartelmann:2010fz}
\begin{align}
  \mu_\pm = \frac{y^2 + 2}{2 y \sqrt{y^2 + 4}} \pm \frac{1}{2} \,.
  \label{eq:magnification-geom-1}
\end{align}
Their interference leads to the observed image, which, for a
point-like source and lens, 
is magnified by a factor
\begin{align}
  \mu = \frac{y^2 +2}{y \sqrt{y^2 + 4}} + \frac{2}{y \sqrt{y^2 + 4}}
        \sin \left(\Omega
               \left[ \frac{y \sqrt{y^2 + 4}}{2}
                          + \log \left| \frac{y + \sqrt{y^2 + 4}}{y - \sqrt{y^2 + 4}} \right|
               \right] \right) \,
  \label{eq:magnification-geom-2}
\end{align}
compared to the unlensed source.  Here, the frequency dependence that
creates the interference pattern enters through the dimensionless
parameter
\begin{align}
  \Omega \equiv 4 G M (1 + Z_L ) \, \omega \,.
  \label{eq:Omega_def_gen}
\end{align}
The requirement $\omega \, \Delta t \gtrsim \mathcal{O}(1)$ 
for the observability
of lensing-induced interference patterns immediately tells us the lens masses
to which a given search will be sensitive.  
For typical values of $\theta, \beta
\sim \theta_E$, the two terms in parentheses in \cref{eq:deltatgeom} are of
similar order, and using \cref{eq:EinsteinAngle} for the Einstein angle,
we find $\Delta t \sim R_s$, where $R_s$ is the Schwarzschild radius of the
lens.  This explains why femtolensing of GRBs~\cite{Gould:1991td},
where photons have energies of order \SI{10}{keV} to  \SI{1}{MeV}, is most sensitive
to lenses between \SI{e-17}{M_\odot} and \SI{e-14}{M_\odot}.  
For fast radio bursts,
with typical photon energies of order \SI{e-6}{eV} (frequencies of
order \si{GHz}), 
lens masses between \SI{e-4}{M_\odot} and \SI{0.1}{M_\odot} will be relevant.
This mass range extends to higher masses than one might naively expect
thanks to the excellent frequency resolution of radio telescopes, which allows
us to resolve the interference pattern even if $\omega \, \Delta t$ is
significantly 
larger than one.

\begin{figure}
  \centering
  \includegraphics[width=.75\textwidth]{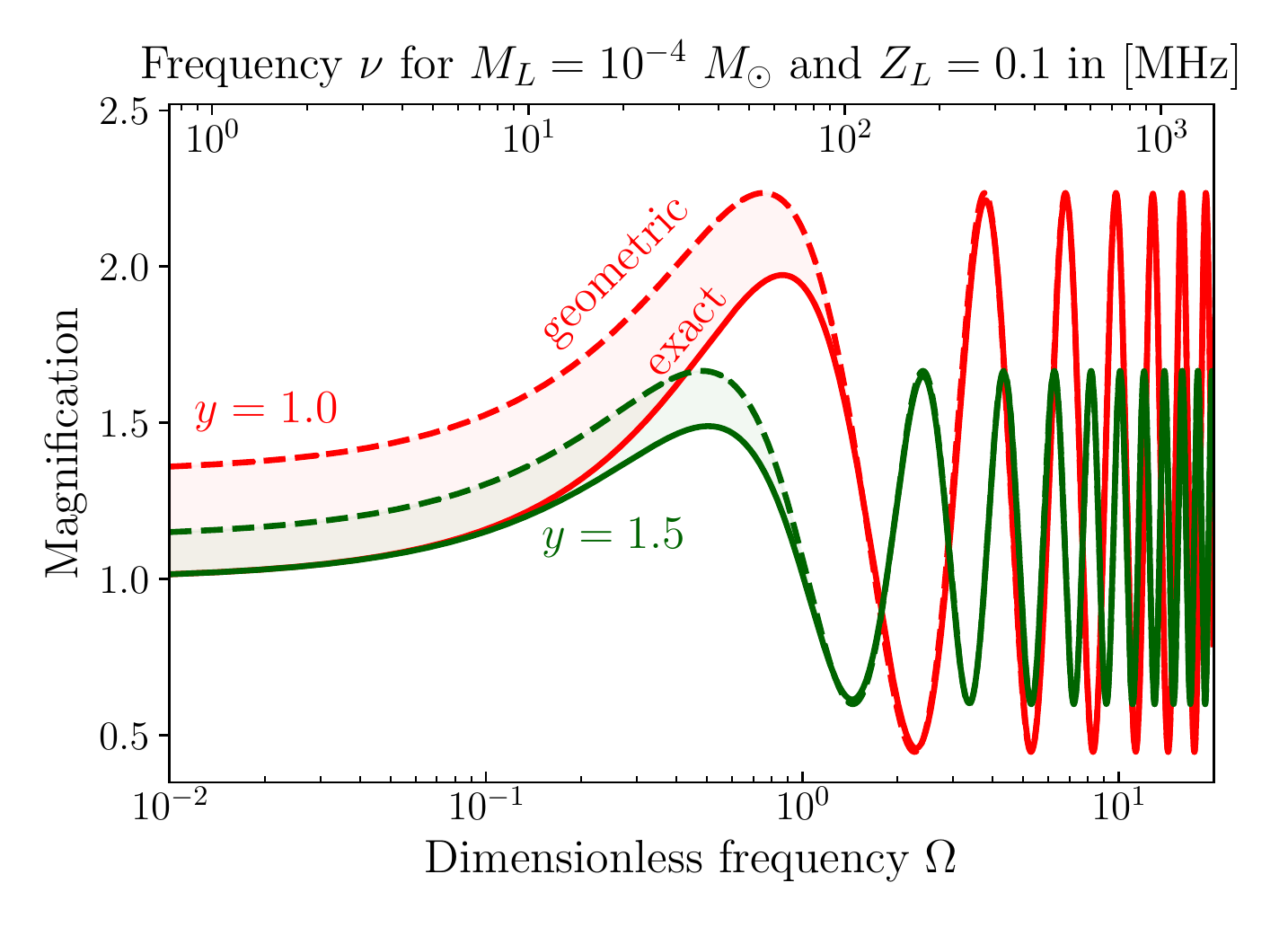}
  \caption{The diffractive lensing signal as a
    function of photon frequency.  The vertical axis displays the
    ratio of the observed 
    intensity with and without lensing, the bottom horizontal axis shows
    the dimensionless frequency defined in \cref{eq:Omega_def_gen}, and the top
    horizontal axis translates $\Omega$ into a physical frequency for a specific
    choice of lens mass and redshift. We compare two different angular
    separations 
    between the lens and the source (red vs.\ green lines), and we also compare
    the geometric optics approximation (dashed) to a full wave optics
    calculation (solid), concluding that, for our purposes ($\nu \sim \si{GHz}$)
    geometric optics is always a valid approximation.}
  \label{fig:femtolensing}
\end{figure}

Let us mention some caveats to the discussion above.  First, we note that at
$\omega \, \Delta t \sim 1$ the geometric optics approximation with two
well-defined lensed images breaks down and wave optics effects need to be taken
into account~\cite{Ulmer:1994ij}.  The observed signal is then found by
evaluating the full Fresnel integral over the lens plane, which leads to the
observed magnification
\begin{align}
  \mu = \bigg| \frac{\Omega}{(2 \pi i) \theta_E^2} 
\int d^2 \vec{\theta}\, e^{i \omega \Delta t(\vec\theta)} \bigg|^2 \,,
  \label{eq:muwave}
\end{align}
where $\Delta t $ is given by \cref{eq:deltatgeom}.  For a point source and a
point-like lens, \cref{eq:muwave} can be evaluated
analytically~\cite{Nakamura:review,Katz:2018zrn}, but in the general case the
integral needs to be performed numerically.  We compare the magnification in
the geometric optics approximation to the result of a full wave optics
calculation in \cref{fig:femtolensing}. As expected, the effect of wave optics
is most pronounced only at small frequencies, $\Omega \lesssim 1$. The typical
frequencies at which we observe FRBs are around \SI{1}{GHz}, which for lens
masses $\gtrsim \SI{e-4}{M_\odot}$ corresponds to $\Omega \gg 1$. We illustrate
in \cref{fig:femtolensing} that in this regime wave optics corrections are
negligible.

A second potential caveat is related to the angular size of the
source. If the latter is too large, but unresolved by the observer,
the integral 
over signals from different regions in the source will wash out any
lensing-induced 
interference patterns.  More precisely, this happens when the argument of the
sine in \cref{eq:magnification-geom-2} varies by an amount $\gtrsim 1$
over the angular diameter of the source. Denoting this diameter,
normalized to $\theta_E$,
as $\sigma_y$, interference fringes are thus only observable if
\begin{align}
  \Omega \sigma_y \sqrt{y^2 + 4} \lesssim 1 \,.
  \label{eq:cond-pointlike}
\end{align}
For a $\SI{e-4}{M_\odot}$ lens about half-way between the observer and a source
at \SI{1}{Gpc}, and for $\Omega \sim 10$, this condition implies that wash-out is avoided
for sources with a physical diameter $\lesssim \SI{e13}{cm}$.
This is a conservative estimate in the sense that, for heavier lenses,
the maximum allowed diameter will only increase. FRBs are variable on
short time scales $t_{\rm var}\lesssim \si{msec}$. Allowing for
relativistic expansion of the emitting matter with a bulk Lorentz
factor $\varGamma$, one can estimate the apparent transverse size of
the source as $a_{S}\simeq t_{\rm var}\varGamma$ 
(see e.g. appendix~A of ref.~\cite{Katz:2018zrn}
for a derivation). One concludes that for realistic values of
$\varGamma<10^5$ the size of the FRB source  
is irrelevant
for the lensing signal. 

We are going to see below that this conclusion may change in the
presence of strong scintillation in the FRB host galaxy. The
scintillation increases an \emph{effective size} of the source and thereby
suppresses the lensing pattern. We defer the study of this effect to
the next section.

The above discussion can be generalized to the case of non-point-like
lenses along the lines of ref.~\cite{Katz:2018zrn}. In particular, it
applies to UCMHs as long as their density slope is steep enough
to produce multiple images of the source.\footnote{In the case of a
  power-law UCMH density profile, $\rho\propto r^{-\delta}$, multiple
  images appear if $\delta>1$. }

%==========================================================================
\section{Scintillation}
\label{sec:scintillation}
%==========================================================================

%  !TEX root =  draft.tex
We now turn to the physics of scintillation in interstellar and intergalactic
medium.  Scintillation leads to distortions of the signals from distant FRBs
that can look very similar to the distortions caused by diffractive
gravitational lensing.  To disentangle the two, we must therefore carefully
model the physics of scintillation.

%--------------------------------------------------------------------------
\subsection{Scintillation Primer}
\label{sec:scint-primer}
%--------------------------------------------------------------------------

When the radio waves from a distant FRB pass through the
turbulent ISM and IGM environments, they will suffer refraction and diffraction,
implying that signals can reach the observer along many different trajectories.
Photons traveling along different paths will interfere, leading to chaotic
modulation of the observed FRB frequency spectrum.  The physics here is similar
to the physics of atmospheric scintillation which makes the stars flicker; as
FRBs emit at different wavelength, their signals are not significantly affected
by the Earth's atmosphere, but mainly by the ISM.

In the subsequent discussion, we follow 
refs.~\cite{1992RSPTA.341..151N,2005handbook,Woan:2011}.
Scintillation can be modeled by assuming that the turbulent plasma which is
responsible 
for multi-path propagation is confined to a thin screen. The position of the
screen is chosen to be about half-way through the scintillating medium. For
scintillation in the Milky Way's ISM, this typically means a
scintillation screen 
$\mathcal{O}(\si{kpc})$ away from the observer; scintillation in the IGM can be
described by a scintillation screen at a distance of $\mathcal{O}(\si{Gpc})$,
for example half-way between the source and the observer.
For simplicity, we perform the analysis in Minkowski spacetime, neglecting
redshift effects.

\begin{figure}
\centering
\includegraphics[width=0.6\textwidth]{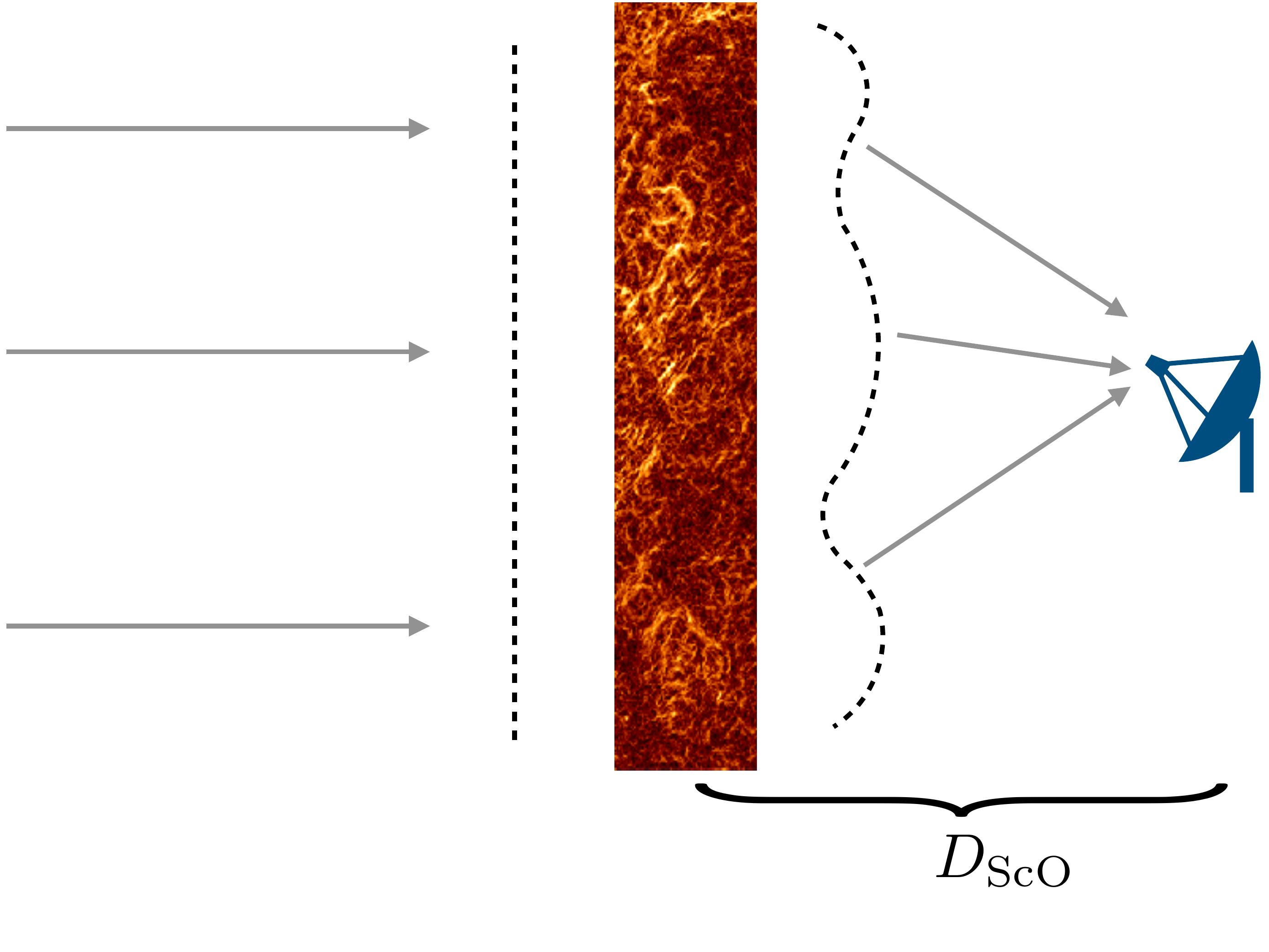}
  \caption{Distortion of a plane wave by scintillating material
    modeled as a thin screen. Dashed lines show the wavefront
    before and after the passage through the screen.}
  \label{fig-screen-1}
\end{figure}

Consider a plane wave of amplitude $f_\text{in}(\omega)$
falling perpendicularly onto a scintillation screen which distorts the
wavefront by adding to it a random phase $\varphi(\omega,\vec{x})$, 
where $\vec{x}$ is the coordinate
on the screen, see \cref{fig-screen-1}.
The signal which an observer at a point $O$, located at a
distance $D_\text{ScO}$ from the screen, receives is given by the Fresnel
integral,
\begin{align}
  f_\text{obs}(\omega) = \frac{\omega f_\text{in}(\omega)}{2\pi i D_\text{ScO}}
                         \int \! d^2x \, e^{i \Phi(\omega, \vec{x})} \,.
  \label{eq:signalO}
\end{align}
with the phase
\begin{align}
  \Phi(\omega, \vec{x}) = \frac{\omega \, x^2}{2 D_\text{ScO}} 
+ \varphi(\omega,\vec{x}) \,.
  \label{eq:phi}
\end{align}
Here, the first term accounts for the geometric phase, which arises because of
the different distances traveled by photons passing the screen at different
positions $\vec{x}$.  The phase perturbation $\varphi(\omega, \vec{x})$ in the
second term is characterized by the \emph{diffractive scale} $r_\text{diff}$,
which is the distance scale on the scintillation screen over which
$\varphi(\omega, \vec{x})$ varies by $\mathcal{O}(1)$. More precisely,
$r_\text{diff}$ is defined using the phase {\it structure function}, 
\begin{align}
  \xi(\omega,|\Delta \vec{x}|)\equiv \ev{\big(\varphi(\omega, \vec{x})
          - \varphi(\omega, \vec{x} + \Delta\vec{x}) \big)^2}
  \label{eq:r-diff-1}
\end{align}
through the equation
\begin{align}
  \xi(\omega,r_\text{diff}) = 1 \,.
  \label{eq:r-diff}
\end{align}
The averaging in \cref{eq:r-diff-1} is performed over an ensemble of
realizations of the random phase
$\varphi(\omega,\vec{x})$. Due to statistical homogeneity of the
scintillation screen, this is equivalent to the
spatial averaging over the screen. 
The diffractive scale derives its name from the fact that it sets the angle
$\theta_\text{diff}$ over which radiation is typically diffracted by the
screen:
\begin{align}
  \theta_\text{diff} = \frac{1}{\omega \, r_\text{diff}} \,.
  \label{eq:theta-diff}
\end{align}
As we will see below, $r_\text{diff}$ is frequency-dependent.

In the case of weak scintillation, where the scintillation phase
$\varphi(\omega, \vec{x})$ varies more slowly with $\vec{x}$ than the geometric
phase,  the radiation the observer receives from a given angular position on
the sky comes from the first Fresnel zone of the screen, namely a region of
size
\begin{align}
  r_F = \sqrt{D_\text{ScO} / \omega} \,.
  \label{eq:rF}
\end{align}
This is the distance over which the geometric phase changes by
$\mathcal{O}(1)$, so the weak scintillation regime is characterized by the
condition $r_\text{diff} \gg r_F$.  In the opposite limit of strong
scintillation ($r_\text{diff} \ll r_F$), the observed signal comes instead from
a region of size
\begin{align}
  r_\text{ref} = \theta_\text{diff} \, D_\text{ScO}
               = \frac{D_\text{ScO}}{\omega \, r_\text{diff}} \,.
  \label{eq:r-ref}
\end{align}
$r_\text{ref}$ is called the \emph{refractive scale}. The different length
scales on the scintillation screen are related by
$r_F = \sqrt{r_\text{diff} \, r_\text{ref}}$.

In this paper we are only interested in the energy spectra of FRBs, not in the
spatial or temporal variation of their intensity. Still, it is instructive to
consider the radiation intensity as a function of position in the observer
plane to understand the designation ``refractive scale'' for $r_\text{ref}$ and
``diffractive scale'' for $r_\text{diff}$~\cite{Moniez:2003ij}.  As the
scintillating medium moves, the picture in the observer plane will move as well
leading to temporal variations in the observed signal.  In the case of strong
scintillation, we expect to see a diffraction pattern of intensity variations
over scales of order $r_\text{diff}$ (``diffractive scintillation''), because
this is by definition the scale over which $\varphi(\omega, \vec{x})$ changes
by $\mathcal{O}(1)$.  We also expect to see intensity variation over larger
scales of order $r_\text{ref}$ (``refractive scintillation'').  As discussed
above, each point on the screen emits radiation predominantly into a cone of
the angular size $\theta_\text{diff}=r_\text{ref}/D_{\text{ScO}}$.  Depending
on the gradient of the refractive index, the cones from neighboring points will
be refracted towards each other, which leads to an increase in observed
radiation intensity, or away from each other, which leads to a decrease in
observed radiation intensity.  These intensity variations are only observable
at scales larger than the cone size, that is $\gtrsim r_\text{ref}$.  They
correspond to temporal variations on timescales $\sim r_\text{ref}/v_\perp$,
where $v_\perp$ is the transverse velocity of the scintillating medium.  For
the study of FRBs, refractive scintillation is not relevant as the displacement
of the screen over the duration of the burst is negligible.

For weak scintillation, the observer will only register intensity variations on
the scale $r_F$. Lines of sight separated by larger scales will effectively
receive radiation from disjoint regions on the screen.

%--------------------------------------------------------------------------
\subsection{Separating Lensing and Scintillation}
\label{sec:separation}
%--------------------------------------------------------------------------

\begin{figure}
\centering
  \includegraphics[width=0.75\textwidth]{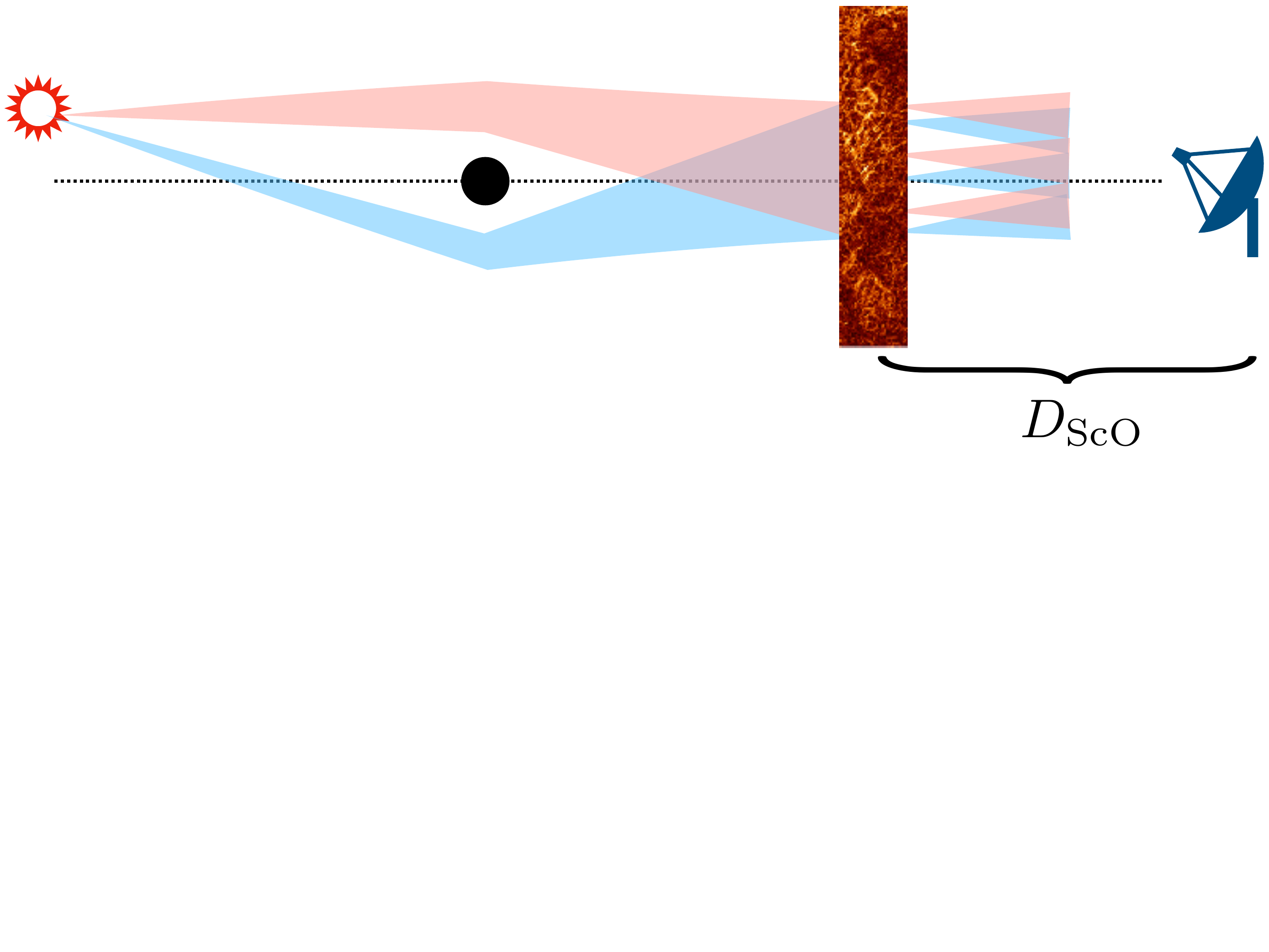}
  \vspace{-4.5cm}
  \caption{A fast radio burst gravitationally lensed by a compact
    object. The superposition of the signals corresponding to the two images
    passes through a scintillating medium before reaching the observer.}
  \label{fig-screen-2}
\end{figure}

Let us now assume that the signal from a distant FRB is gravitationally lensed
by a compact object such as a PBH close to the line of sight before it is
distorted by a scintillation screen (see illustration in \cref{fig-screen-2}).
For the purposes of this  discussion we will restrict ourselves to lensing in
the geometric optics limits. We have already seen in \cref{sec:femto} that this
approximation is sufficient for the lens masses we are interested in.

We describe the time-dependent FRB signal at the source as
\begin{align}
  F_\text{in}(t) = \int \! \frac{d\omega}{2\pi} \, f_\text{in}(\omega) e^{-i \omega t} \,,
\end{align}
where $f_\text{in}(\omega)$ is the amplitude of the radiation at
frequency $\omega$.
Lensing in the geometric optics approximation splits this signal into two
parts with different amplitudes $A$ and $B$, and with a relative propagation
time delay $\Delta t$ (see \cref{eq:deltatgeom}):
\begin{align}
  F_\text{lens}(t) = A F_\text{in}(t) + B F_\text{in}(t - \Delta t) \,.
  \label{eq:split}
\end{align}
A scintillation screen between the lens and the observer additionally
imparts random noise on the lensed signal.
The observed signal from a distant source, after passing a gravitational
lens (e.g.\ a PBH) and then a scintillation screen
(e.g.\ due to the 
Milky Way's ISM), can be written as
\begin{align}
  f_\text{obs}(\omega) = \frac{\omega f_\text{in}(\omega)}{2\pi i D_\text{ScO}}
                         \int \! d^2x \, e^{i \Phi(\omega, \vec{x})}
                         \big(
                           A e^{i\omega \delta t_A(\vec{x})}
                         + B e^{i\omega [\Delta t + \delta t_B(\vec{x})]}
                         \big) \,.
  \label{eq:sig-coh}
\end{align}
Here, the terms $\delta t_{A,B}(\vec{x})$ account for the variation of the
lensing-induced time delays with $\vec{x}$.  This variation originates from the
fact that the lens is seen under a different angle relative to the source from
different points on the screen.
Taking the square of this expression we obtain the intensity spectrum,
\begin{align}
 |f_\text{obs}(\omega)|^2= |f_\text{in}(\omega)|^2
\Big(|{\cal A}(\omega)|^2+|{\cal B}(\omega)|^2
+{\cal A}^*(\omega){\cal B}(\omega)\, e^{i\omega\Delta t}
+{\cal A}(\omega){\cal B}^*(\omega)\, e^{-i\omega\Delta t}
\Big)\;,
\label{eq:transfer}
\end{align}
where
\begin{align}
{\cal A}(\omega)=\frac{A\,\omega}{2\pi i D_\text{ScO}}\int d^2x\,
e^{i(\Phi(\omega, \vec{x})+\omega \delta t_A(\vec{x}))}
\label{eq:amplA}
\end{align}
and similarly for ${\cal B}(\omega)$. We observe the presence of
interference terms characteristic of diffractive lensing. If the
amplitudes ${\cal A}$, ${\cal B}$ were constant, these terms would
lead to regular periodic modulation of the observed intensity, as
discussed in \cref{sec:femto}. Scintillation complicates the picture by
making the amplitudes frequency dependent. Nevertheless, the lensing
signal can still be extracted if the product of the amplitudes 
${\cal  A}^*(\omega){\cal B}(\omega)$ contains a slowly varying
component that survives upon averaging over frequency intervals
$\Delta \omega\gtrsim (\Delta t)^{-1}$. Indeed, in this case the
Fourier transform of the spectrum (\ref{eq:transfer}) will have a peak
coresponding to the lensing time delay $\Delta t$.  

To estimate the size of the slowly varying component we consider the
product of the amplitudes smeared with a Gaussian function of width
$\Delta\omega$, 
\begin{align}
  Q(\omega,\Delta\omega)=\int\!\frac{d\omega'}{\sqrt{2\pi}\Delta\omega}
    \,e^{-\frac{(\omega'-\omega)^2}{2\Delta\omega^2}} 
    {\cal A}^*(\omega') {\cal B}(\omega')\;.
  \label{eq:Qvar}
\end{align}
Its characteristic value is given by 
the ensemble average $\langle |Q(\omega,\Delta\omega)|^2\rangle$
over random realizations of the scintillation phase. This average
involves the correlator of four amplitudes,
\begin{align}
  \langle {\cal A}^*(\omega'){\cal B}(\omega'){\cal A}(\omega''){\cal B}^*(\omega'')\rangle\;.
  \label{eq:4point}
\end{align}
 In the case of strong
scintillation the integral (\ref{eq:amplA}) extends over many Fresnel
regions. The contributions of different regions are essentially
statistically independent. Then, by the central limiting theorem, the
statistics of the amplitudes ${\cal A}(\omega)$ and ${\cal B}(\omega)$
are close to Gaussian and we can replace (\ref{eq:4point}) by
\begin{align}
  \langle {\cal A}^*(\omega'){\cal A}(\omega'')\rangle
  \,
  \langle {\cal B}(\omega'){\cal B}^*(\omega'')\rangle
  +\langle {\cal A}^*(\omega'){\cal B}(\omega')\rangle
  \,
  \langle {\cal A}(\omega''){\cal B}^*(\omega'')\rangle\;,
  \label{eq:Wick}
\end{align}
where we have used that the correlator $\langle {\cal A}(\omega'){\cal
B}(\omega'')\rangle$ vanishes.  Our next task is to estimate the two-point
correlators entering into this expression.

Let us start with 
\begin{align}
\langle {\cal A}^*(\omega') {\cal A}(\omega'')\rangle
\equiv A^2 {\cal C}_0(\omega''-\omega')\;,
\label{eq:autocor0}
\end{align}
where on the r.h.s. we introduced the frequency autocorrelation
function for the signal coming from a single image. 
A similar equation holds for the correlator of the ${\cal B}$
amplitudes. 
An important characteristic of the scintillation-induced noise is its
decorrelation bandwidth $\omega_\text{dec}$. 
The noise amplitude is correlated at
frequencies separated by less than $\omega_\text{dec}$ 
and uncorrelated otherwise.
An estimate for $\omega_\text{dec}$ is obtained by recalling 
that the observer
sees a patch of diameter $r_\text{ref}$ on the scintillation screen.
Over this distance, the geometric phase (first term in \cref{eq:phi}) varies
by an amount $\omega r_\text{ref}^2 / (2 D_\text{ScO})$. A shift in
frequency by $2D_\text{ScO}/r^2_\text{ref}$ 
changes the phases at different
points of the patch by order unity resulting in decorrelation of the
scintillation noise. 
Thus we arrive at the estimate
\begin{align}
  \omega_\text{dec} \sim \frac{2 D_\text{ScO}}{r_\text{ref}^2} \,.
  \label{eq:Omega-dc}
\end{align}
Note that we assumed above that the random 
phase
$\varphi(\omega, \vec{x})$ (second term in \cref{eq:phi}) changes with
frequency in the same way or more slowly than the geometric phase. This is
indeed the case for scintillation in ionized plasma, where $\varphi$
is inversely proportional to frequency (see the next subsection). 
The decorrelation bandwidth sets the range of the autocorrelation
function (\ref{eq:autocor0}): one expects that ${\cal C}_0$
is of order 1 if $|\omega''-\omega'|\lesssim \omega_\text{dec}$ and
quickly decreases outside this range. 

The second term in \cref{eq:Wick} contains the cross-correlation
between the amplitudes of the two lensed images at fixed frequency
$\langle {\cal A}^*(\omega){\cal B}(\omega)\rangle$.
Its magnitude depends on whether the two images are distorted by
scintillation screen in the same way (coherently) or independently
(incoherently). In the first case the correlation is order-one, whereas
in the second case it is suppressed. To determine the conditions for
coherent/incoherent distortion, let us consider the variation 
$\delta t_B(\vec{x})-\delta t_A(\vec{x})$ of the lensing-induced time
delay over the scintillation screen. This can be found by noting that
a change of position on the screen changes the difference between the
viewing angles of the lens and the source by $\delta\beta=(x
D_{LS})/(D_L D_S)$. Hence, the variation of the time delay is given by
\begin{align}
\delta t_B-\delta t_A=\frac{d\Delta t}{d\vec{\beta}}\cdot 
\frac{\vec{x}\, D_{LS}}{D_L D_S}=\Delta\vec{\theta}\,\vec{x}\;,
\end{align}  
where $\Delta\vec\theta$ is the angular distance between the $A$ and
$B$ images. In the second equality we have used \cref{eq:deltatgeom}. Due
to scintillation, the observer receives photons from a region of
size $r_\text{ref}$ on the screen. If the variation of the
lensing-induced phase over this region is small, we can neglect it in
the integral (\ref{eq:amplA}) for the amplitudes and obtain,
\begin{align}
\langle{\cal A}^*(\omega){\cal B}(\omega)\rangle
=AB\bigg\langle \bigg|\frac{\omega}{2\pi i D_\text{ScO}}
\int d^2x\, e^{i\Phi(\omega,\vec{x})}\bigg|^2
\bigg\rangle\sim AB\;.
\label{eq:corcasec}
\end{align} 
In the opposite regime a fast lensing-phase variation leads to strong
suppression of the correlation between the amplitudes. Thus, in
general we can write,
\begin{align}
\langle{\cal A}^*(\omega){\cal B}(\omega)\rangle=AB\cdot {\cal
  U}(\omega\,\Delta \theta\, r_\text{ref})\,,
\label{eq:ABcorU}
\end{align}
where the function ${\cal U}$ is of order unity when its argument is
less than 1 and quickly vanishes outside this range.\footnote{Under
  general assumptions about the statistical properties of the
  scintillation phase, one can show that ${\cal U}(z)$ is
  exponentially suppressed at $z>1$.}

We now return to the smeared amplitude of the interference term,
\cref{eq:Qvar}. Collecting our
previous results from \cref{eq:autocor0,eq:ABcorU} we obtain,
\begin{align}
\langle |Q(\omega,\Delta\omega)|^2\rangle
= A^2 B^2\bigg\{\int\frac{d\omega_-}{2\sqrt{\pi}\Delta\omega}
\,e^{-\frac{\omega_-^2}{4\Delta\omega^2}}\big|{\cal C}_0(\omega_-)\big|^2
+\big|\bar {\cal U}(\omega\,\Delta\theta\,r_\text{ref})\big|^2\bigg\}\;,
\label{eq:Qvar2}
\end{align}
where we have denoted by $\bar{\cal U}$ the correlator
(\ref{eq:ABcorU}) averaged over a range of frequencies
$\Delta\omega$. Extraction of the lensing signal is possible whenever
either of the two terms in brackets is sizable, implying that the interference
amplitude has a piece that does not vanish when averaged over frequency
intervals $\Delta\omega \gtrsim (\Delta t)^{-1}$. If $\Delta\omega$ is
smaller than the decorrelation bandwidth, the autocorrelation function
in the first term can be replaced by unity and the whole integral
equals 1. In the opposite regime $\Delta\omega\gg \omega_\text{dec}$
it becomes suppressed as $\omega_\text{dec}/\Delta\omega$. 
Recalling that we want the averaging
interval $\Delta\omega$ to be at least as large as $(\Delta t)^{-1}$,
we conclude that the first term in \cref{eq:Qvar2} is $\mathcal{O}(1)$ for
long lensing delays $\Delta t\gg (\omega_\text{dec})^{-1}$. The magnitude
of the second term in \cref{eq:Qvar2} is controlled by the angular
separation between the lensed images. Given that the characteristic
size of $\Delta\theta$ is set by the Einstein angle $\theta_E$, we can
rephrase the condition for this term to be significante as
$\omega\,\theta_E\, r_\text{ref}<1$. In this way we arrive at four
possible scenarios, which are illustrated in \cref{fig:scint-regions}: 
\begin{enumerate}
  \item[{\bf (a)}] \underline{$\Delta t \gg (\omega_\text{dec})^{-1}$, 
    ~$\omega\,\theta_E\,r_\text{ref} \ll 1$}
    (region above the red and blue lines in \cref{fig:scint-regions}).
    In this case, both terms in \cref{eq:Qvar2} are sizeable.
    The lensing phase $\omega \Delta t$
    varies much faster with frequency 
    than the scintillation phase $\Phi(\omega, \vec{x})$, and
    does not change substantially across the scintillation screen.
    Therefore, a clear lensing pattern can be observed in the form of rapid
    periodic variations of the radiation intensity with $\omega$.
    The envelope of the lensing wiggles in the frequency spectrum
    is modulated over frequency intervals $\gg (\Delta t)^{-1}$ by scintillation
    effects and by the variation of the lensing phase across the screen.

    This behavior is clearly visible in the top panels of \cref{fig:3cases}.
    In the left part of this figure, we show the transfer function
    \begin{align}
      |T(\nu)|^2 \equiv \bigg|
      \frac{f_\text{obs}(\nu)}{f_\text{in}(\nu)} \bigg|^2 \,, 
      \label{eq:transfer-function}
    \end{align}
    while in the middle part, we plot the autocorrelation function of the
    normalized signal amplitude,
    \begin{align}
      {\cal C}(d\nu) \equiv \frac{\big|\!\ev{T^*(\nu) \, T(\nu + d\nu)}
        \!\big|}{\ev{|T(\nu)|^2}} \,. 
      \label{eq:autocorr}
    \end{align}
    Here $\nu = \omega / (2\pi)$ is the radiation frequency and
    averaging is performed over a large frequency interval.  In the right
    part of \cref{fig:3cases}, we show the Fourier transform of $|T(\nu)|^2$.
    As expected, the lensing signal is clearly visible, with its envelope only
    marginally modulated due to scintillation.

\begin{figure}
  \centering
  \includegraphics[width=0.6\textwidth]{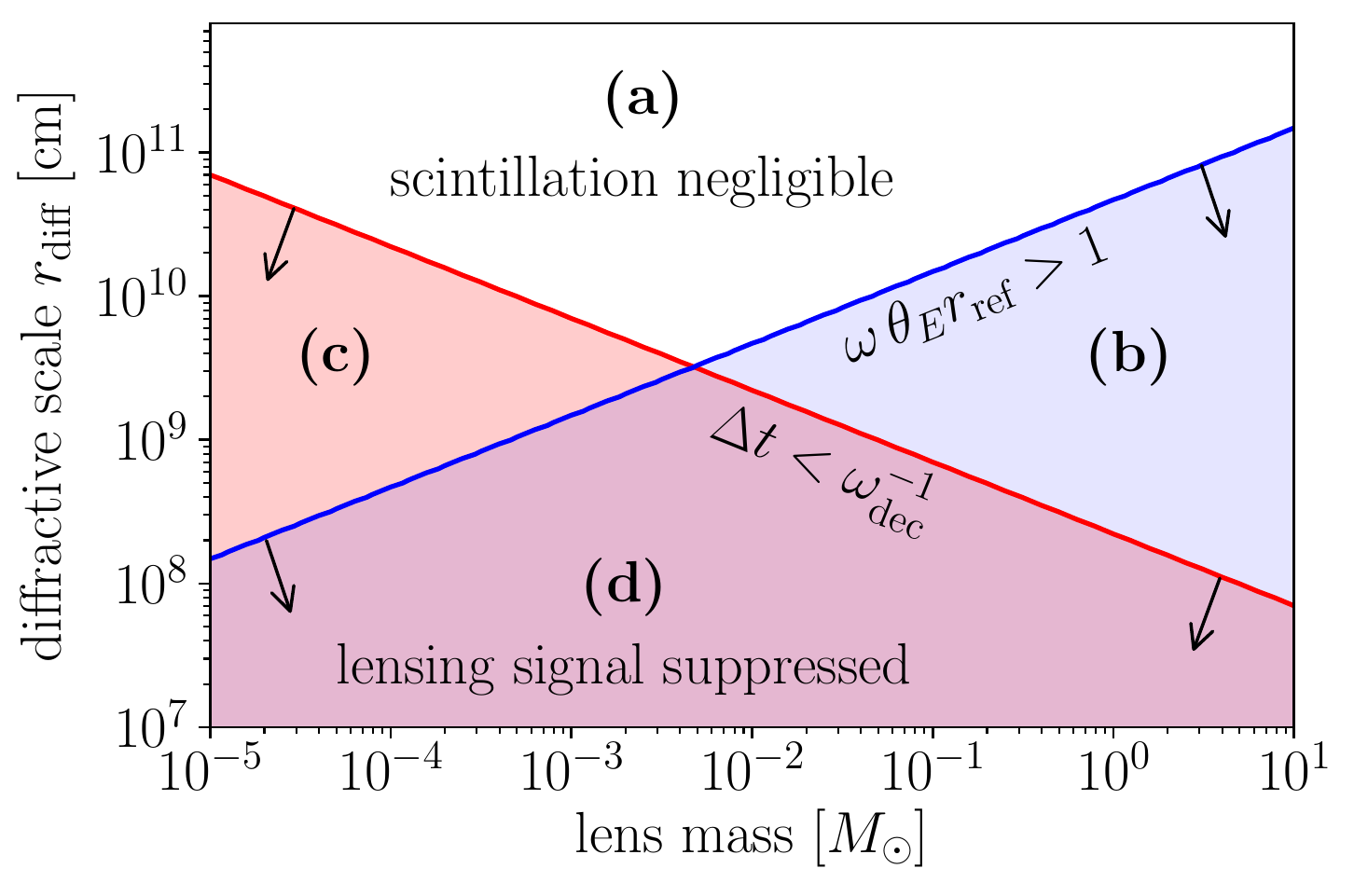}
  \caption{Observability of FRB lensing by compact dark matter 
    objects as a function of
    the lens mass $M$ and of the diffractive scale corresponding to
    interstellar scintillation in the Milky Way. We assume an FRB at a
    comoving distance of \SI{1}{Gpc}, and a lens at \SI{0.5}{Gpc}. The
    angular separation between the lens and the (unlensed) source,
    normalized to the Einstein angle, is 
 $y=\beta/\theta_E=0.5$. We consider
    $\SI{1}{GHz}$ radio waves, which are perturbed by a scintillation screen
    at \SI{1}{kpc} form the observer.}
  \label{fig:scint-regions}
\end{figure}

\begin{figure}
  \centering
  \includegraphics[width=\textwidth]{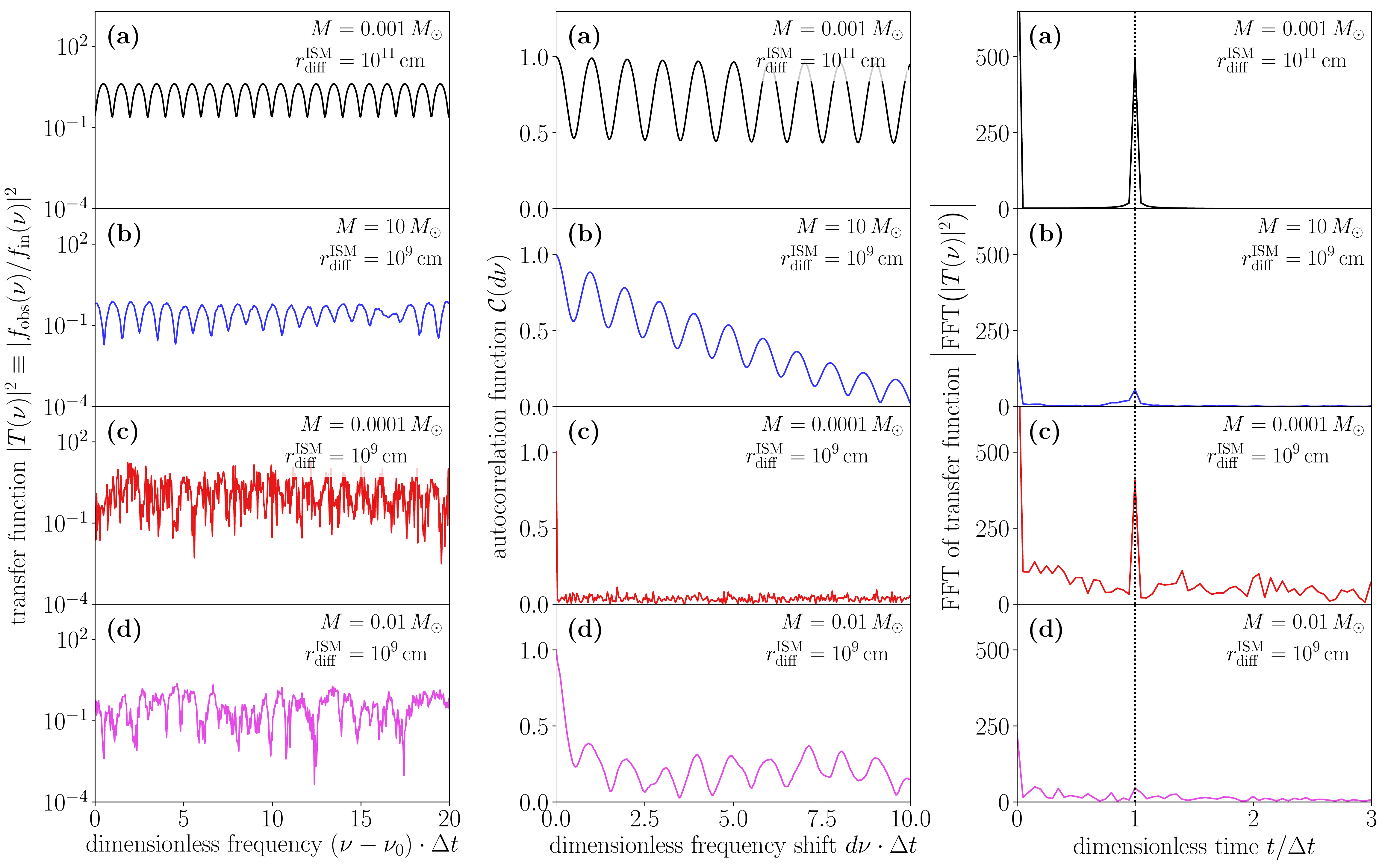}
  \caption{From top to bottom, we illustrate the four qualitatively different
    scintillation/lensing regimes discussed in the main text. The
    panels on the left show in each case the transfer function $|T(\nu)|^2$,
    that is the observed radiation intensity, normalized to the radiation
    intensity without lensing and scintillation (see
    \cref{eq:transfer-function}).  In the middle panels we plot the
    autocorrelation function, \cref{eq:autocorr}. The panels on the right
    contain the Fourier transform of $|T(\nu)|^2$, with the lensing peak
    highlighted by a vertical dotted line. In all panels, the horizontal axis
    is normalized in terms of the lensing time delay $\Delta t$, and we use
    $\nu_0 = \SI{1}{GHz}$.  Values of $r_\text{diff}$ are quoted at the
    midpoint of the spectrum.  The plots have been produced using the
    simulation code described in \cref{sec:simulation-and-results}
    \cite{github}, assuming an interstellar scintillation screen at
    \SI{1}{kpc}, and a lens at $D_L = D_S/2$, $y = \beta/\theta_E = 0.5$.}
  \label{fig:3cases}
\end{figure}

  \item[{\bf (b)}] 
    \underline{$\Delta t \gg (\omega_\text{dec})^{-1}$, 
    ~$\omega\,\theta_E\,r_\text{ref}\gg 1$}
    (region below the blue, but above the red line in
    \cref{fig:scint-regions}).
    The variation of the lensing phase over the scintillation screen is fast,
    so the two lensed images are distorted incoherently. Thus, the second term
    in \cref{eq:Qvar2} is suppressed.
    Nevertheless, the first term remains of order unity as the
    amplitudes ${\cal A}(\omega)$ and ${\cal B}(\omega)$ vary slowly as
    functions of $\omega$. A clear lensing signal still survives both
    in the transfer function and in the autocorrelation function. This is
    illustrated in the second row of panels in \cref{fig:3cases}.

  \item[{\bf (c)}] 
    \underline{$\Delta t \ll (\omega_\text{dec})^{-1}$, 
    ~$\omega\,\theta_E\,r_\text{ref} \ll 1$}
    (region below the red, but above the blue line in
    \cref{fig:scint-regions}).
    When the time delay $\Delta t$ between the two lensed images is shorter
    than the inverse of the decorrelation bandwidth $\omega_\text{dec}$,
    the scintillation factor $e^{i \Phi(\omega, \vec{x})}$ in \cref{eq:sig-coh}
    varies much faster than the lensing factor $e^{i\omega \Delta t}$. The first
    term in \cref{eq:Qvar2} is then small.
    On the other hand, we can neglect 
    the variation of the lensing phase over the screen, $\omega
    \, \delta t(\vec{x})$, so the two lensed images are distorted
    coherently and the second term in \cref{eq:Qvar} is sizable.
    The lensing signal will be discernible as a modulation
    of the envelope of an otherwise chaotic spectrum. 
    Indeed, the panels in the third row in 
    \cref{fig:3cases} illustrate this behavior: we
    observe high-frequency scintillation noise, superimposed on 
    regular periodic oscillations 
     due to lensing.
    As expected, the signal disappears from the autocorrelation function,
    but remains in the Fourier transform of the transfer function.

  \item[{\bf (d)}] 
    \underline{$\Delta t \ll (\omega_\text{dec})^{-1}$, 
    ~$\omega\,\theta_E\,r_\text{ref}\gg 1$}
    (region below the red and blue lines in \cref{fig:scint-regions}).
    In this case, the fast variations of the scintillation factor
    $e^{i \Phi(\omega, \vec{x})}$ combines with incoherent distortion
    of the images. The amplitude ${\cal A}^*(\omega){\cal B}(\omega)$
    of the interference term in \cref{eq:transfer} does not contain a
    slowly varying component and the lensing signal is strongly
    suppressed, see the last row in \cref{fig:3cases}.
\end{enumerate}
We conclude that a lensing signal is in principle observable in regimes
(a), (b) and (c), but strongly suppressed or completely unobservable 
 in regime (d).

However, there are practical considerations that may render the signal
unobservable even in regimes (a), (b), (c).  The distance between
subsequent lensing fringes in the spectrum, $(\Delta t)^{-1}$, should
be significantly larger than the instrumental frequency resolution,
but significantly smaller than the instrumental bandwidth.  The first condition
is violated for too large lens masses, the second one for lens masses that are
too small.\footnote{Note that at masses $\gtrsim 10 M_{\odot}$ one would expect
lensing time delays $\gtrsim \SI{e-3}{sec}$, comparable to the typical duration
of FRBs.  This would correspond to modulation on \si{kHz} scales in the frequency
spectrum.  However, in this regime, it seems more convenient to work directly
with the temporal profile of the FRBs~\cite{Dai:2017twh}.}  
This effectively restricts the search for compact objects
using diffractive lensing of 
FRBs to the mass range between $\sim \SI{e-4}{M_\odot}$
and $\sim \SI{0.1}{M_\odot}$.

\begin{figure}
\centering
\includegraphics[width=0.75\textwidth]{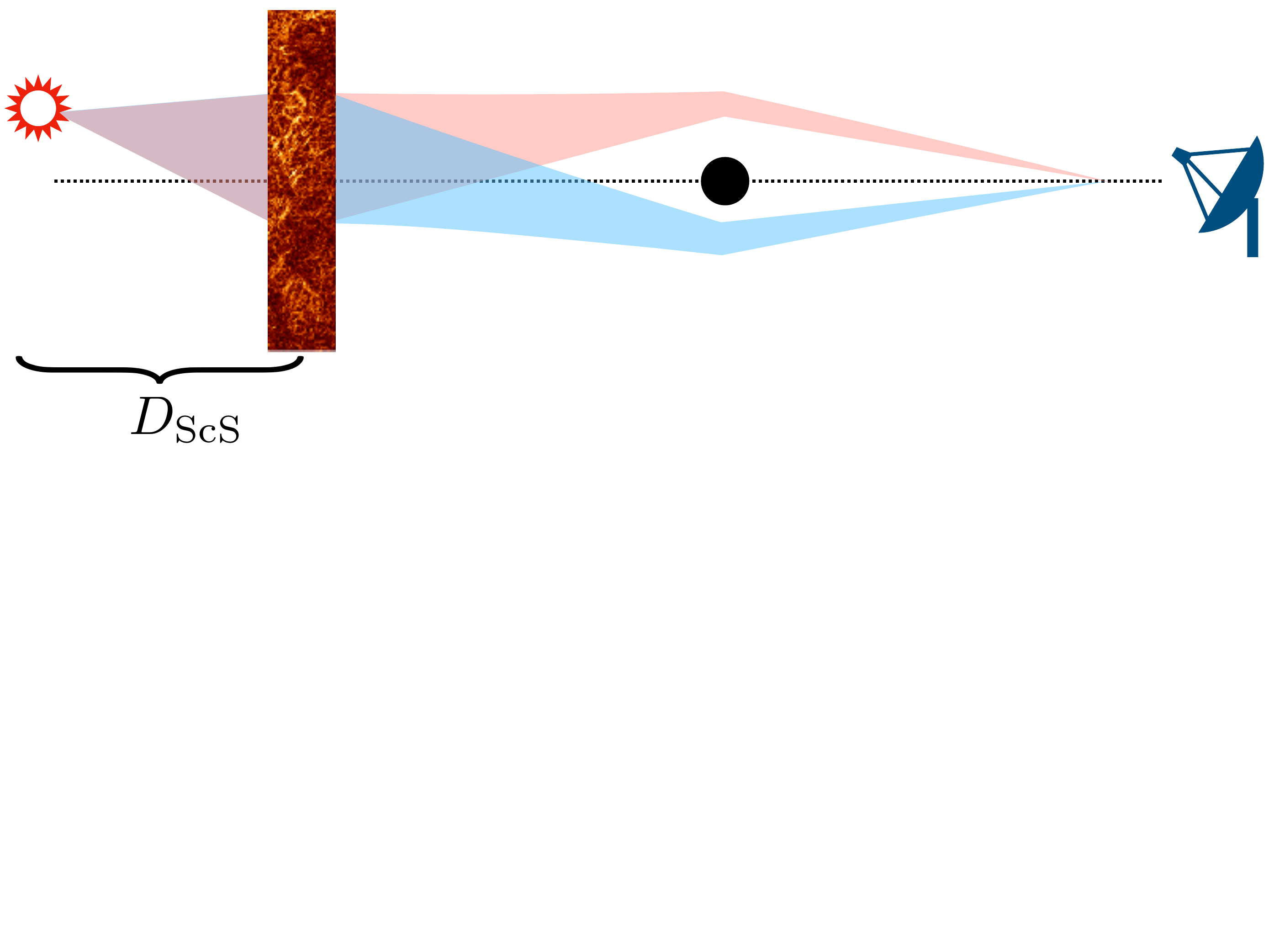}
\vspace{-4.5cm}
  \caption{Setup with the scintillation screen between the source
    and the lens.}
  \label{fig-screen-3}
\end{figure}

The above discussion applies also to the case of a scintillation
screen placed between the source and the lens, see
\cref{fig-screen-3}. This setup describes scintillation in the ISM of
the FRB host galaxy or in the IGM between the lens and the host. The
amplitude of the observed signal is again given by \cref{eq:sig-coh},
with the screen--observer distance $D_\text{ScO}$ replaced by the
screen--source distance $D_\text{ScS}$. The rest of the analysis
proceeds without change. Notice that strong scintillation effectively
spreads the source into a patch of radius $r_\text{ref}$ on the
scintillation screen. The condition $\omega\,\theta_E\,r_\text{ref}<1$
for coherent distortion of the lensed images can then be reinterpreted
as the restriction (\ref{eq:cond-pointlike}) on the size of the source
required for diffractive lensing. One may wonder why one can
discern the lensing pattern in regime (b) even
if this condition is violated. The reason is 
that the signal from each point on the
lensing screen is correlated at different frequencies within a finite
decorrelation bandwidth, which is not the case for a truly incoherent
source.  

%--------------------------------------------------------------------------
\subsection{The Turbulent Interstellar and Intergalatic Medium}
\label{sec:ism-igm}
%--------------------------------------------------------------------------

For a quantitative description of scintillation and its effect on FRB signals,
we need to discuss in more detail the properties of the interstellar
medium (ISM) and 
intergalactic medium (IGM).  We focus in particular on interstellar
and intergalactic plasma, neglecting the neutral gas.
The reason is that the refractive index of a plasma deviates from
unity by much more than the refractive index of neutral gas, so that its impact
on the phase fluctuations that cause scintillation is much larger.

\subsubsection{The Interstellar Medium}
%--------------------------------------

Propagation in plasma modifies the dispersion relation of
electromagnetic waves with wavenumber $k$,
\begin{align}
\omega^2=k^2+\omega_p^2\;.
 \label{eq:dispersion}
\end{align}
Here the plasma frequency is expressed in terms of the electron number
density $n_e$, the electron mass $m_e$ and the electromagnetic fine
structure constant $\alpha$, 
\begin{align}
  \omega_p = \sqrt{\frac{4 \pi \alpha n_e}{m_e}} \,.
  \label{eq:omega-p}
\end{align}
As a consequence, a wave propagating in plasma acquires a frequency
dependent phase shift,
\begin{align}
\varphi(\vec{x}) = -\frac{2\pi \alpha}{m_e\omega}\int n_e(\vec{x}, z) \, dz\;,
\label{eq:phaseshift}
\end{align}
with the integral taken along the line of sight. The phase shift has
strong frequency dependence and causes a time delay between signals in
different frequency bands. In practice, the net time delay due to the average
electron density $\bar n_e$ is removed by the dedispersion
procedure that is typically applied to FRB data.
Scintillation is due to a residual phase shift generated by
random fluctuations in the electron density $\delta n_e=n_e-\bar n_e$.  

We assume that on the scales we are interested in the density
fluctuations obey statistical homogeniety and isotropy. Hence they are
characterized by an isotropic power spectrum related to the density
correlation function as,\footnote{We use the notation $\kappa$ for the
wavenumber of the fluctuations to distinguish it from the wavenumber
of the electromagnetic wave $k$.}
\begin{align}
\langle \delta n_e(\vec{X}+\Delta\vec{X})\,\delta n_e(\vec{X})\rangle
=\int d^3\kappa\, e^{i\vec{\kappa}\Delta\vec{X}}\, P_n(\kappa)\;,
 \label{eq:ne-power-spect}
\end{align}
Here, $\vec{X} = (\vec{x}, z)$ is a three-dimensional coordinate vector,
and $\kappa\equiv |\vec{\kappa}|$. Observations indicate
that in a wide range of wavenumbers the power spectrum obeys
Kolmogorov-like scaling characteristic of a turbulent behavior
\cite{1981Natur.291..561A,Armstrong:1995}, 
\begin{align}
  P_n(\kappa) = C_n^2\, \kappa^{-11/3} \,,~~~~~~~~~
\frac{2\pi}{l_{\rm out}}<\kappa<\frac{2\pi}{l_{\rm in}}\;.
  \label{eq:P-Kolmogorov}
\end{align}
The {\it outer} scale of interstellar turbulence is estimated to be $l_{\rm
  out}\simeq 10\div 100\, \text{pc}$ and the {\it inner} scale $l_{\rm in}$ is
smaller than $10^8\,\text{cm}$; the level of turbulence
is \cite{2004ARA&A..42..211E,2004ARA&A..42..275S}
\begin{align}
C_n^2\simeq (10^{-4}\div 10^{-3})\,\text{m}^{-20/3}\;.
\label{eq:turblevel}
\end{align}
The amplitude of fluctuations with length scale $l$ can be
estimated from the power spectrum as
\begin{align}
  \delta n_e\Big|_l\sim
    \bigg(4\pi \int_0^{2\pi/l} \! d\kappa \, \kappa^2 P_n(\kappa)\bigg)^{1/2}
  \sim (6\pi C_n^2)^{1/2}\,\bigg(\frac{l}{2\pi}\bigg)^{1/3}\;. 
  \label{eq:deltan}
\end{align}
Note that the fluctuation amplitude increases with scale and at the
outer scale becomes comparable to the mean electron density in the
Galaxy,
\begin{align}
\delta n_e\Big|_{l_{\rm out}}\sim \bar n_e\;,
\end{align} 
where $\bar n_e=(0.01\div 0.1)\,\text{cm}^{-3}$ \cite{2002astro.ph..7156C}.
The dominant energy source for ISM turbulence 
is most likely supernova remnants, but other sources
like stellar winds or protostellar outflows can also play a role
\cite{2004ARA&A..42..211E,Shukurov:2011}. 
Even though Kolmogorov scaling (\ref{eq:P-Kolmogorov}) 
is widely used for the description of the turbulent
ISM, its origin remains a matter of debate.
The original Kolmogorov theory
\cite{Kolmogorov:1941a,Kolmogorov:1941c}
predicts the spectrum of energy fluctuations in the inertial range of
an incompressible fluid $P_E(\kappa)\propto \kappa^{-11/3}$. It is not
entirely 
clear why the spectrum of density fluctuations in the compressible ISM
should follow the same power-law, for a review of existing proposals
see~\cite{2004ARA&A..42..211E}. Following the common practice, we will
adopt (\ref{eq:P-Kolmogorov}) in our estimates.

We can now express the phase structure function $\xi(\omega,r)$
from \cref{eq:r-diff-1}
in terms of $P_n(\kappa)$ by plugging the relation \eqref{eq:phaseshift}
between $\varphi(\omega, \vec{x})$ and $n_e(\vec{X})$, as well as
the density correlation function \cref{eq:ne-power-spect}, into
\cref{eq:r-diff-1} \cite{doi:10.1146/annurev.aa.28.090190.003021}:
\begin{align}
  \xi(\omega,r)
    &= 8\pi^2 L\bigg(\frac{2\pi\alpha}{m_e\omega}\bigg)^2
       \int_0^\infty d\kappa\,\kappa\,(1-J_0(\kappa r))P_n(\kappa) \,.
  \label{eq:phi-struct}
\end{align}
Here, $L$ is the thickness of the ISM layer traversed by the wave and
$J_0(z)$ is the Bessel function. Substitution of the Kolmogorov
spectrum (\ref{eq:P-Kolmogorov}) yields,
\begin{align}
\xi(\omega,r)=\bigg(\frac{r}{r_\text{diff}(\omega)}\bigg)^{5/3}\;,
\label{eq:phi-struct-2}
\end{align} 
with the diffractive scale
\begin{align}
r_\text{diff}=0.068\times
(C_n^2 L)^{-3/5}\,\bigg(\frac{m_e\omega}{2\pi\alpha}\bigg)^{6/5}
\;.
\label{eq:rdiff1}
\end{align}
For the typical thickness of the galactic disk $L\sim 1\,\text{kpc}$
we obtain,
\begin{align}
r_{\rm diff}=4.2\times
10^9\,\text{cm}\,\bigg(\frac{L}{\text{kpc}}\bigg)^{-3/5}
\bigg(\frac{C_n^2}{10^{-4}\,\text{m}^{-20/3}}\bigg)^{-3/5}
\bigg(\frac{\nu}{\text{GHz}}\bigg)^{6/5}\;.
\label{eq:rdiff2}
\end{align}
According to \cref{fig:scint-regions}, this belongs to the range of
values admitting separation of the lensing effects from scintillation.
Note a rather steep growth of the diffractive scale with frequency
which results in a rapid suppression of scintillation for
frequencies above a few GHz.

\subsubsection{The Intergalactic Medium}
%---------------------------------------

While turbulence and thus scintillation in the ISM is already fraught with large
uncertainties, even less is known about the properties of the IGM.
In fact, much of what we know about IGM turbulence comes from observations
of FRBs \cite{Macquart:2013nba, Cordes:2016rpr, Zhu:2018kwp}. While to the best 
of our knowledge IGM scintillation has not yet been measured experimentally, there are 
indications that it might be accessible to SKA~\cite{Koay:2014gsa}.  

In the absence of direct measurements, we are forced to rely on
theoretical estimates. Assuming Kolmogorov turbulence, we can estimate
the magnitude of IGM density fluctuations, $C_n^2$,
by equating the density fluctuation at the outer
scale to the average IGM electron density $\bar n_e \simeq
\SI{2.2e-7}{cm^{-3}}$, corresponding to fully ionized hydrogen
and helium~\cite{Koay:2014gsa}.\footnote{For simplicity, we
  neglect here the effects of redshift. They can be taken into account
systematically along the lines of ref.~\cite{Macquart:2013nba}.}
Considering the balance between
heating and dissipation of energy in the IGM,
ref.~\cite{2014ApJ...785L..26L} estimated 
the outer scale to be of order $l_{\rm out}\sim 10^{24}\,\text{cm}$
($\sim 0.3\,\text{Mpc}$). This leads to
\begin{align}
  C_n^2\sim \SI{2e-17}{m^{-20/3}} \,
    \bigg(\frac{\bar n_e}{\SI{2.2e-7}{cm^{-3}}}\bigg)^{2}
    \bigg(\frac{l_{\rm out}}{\SI{e24}{cm}}\bigg)^{-2/3}\;.
  \label{eq:CnIGM}
\end{align}
Substituting this into \cref{eq:rdiff2} and normalizing the thickness
of IGM to \SI{1}{Gpc} we obtain the diffractive scale,
\begin{align}
  r_{\rm diff}\sim \SI{4.5e13}{cm}\,\bigg(\frac{L}{\text{Gpc}}\bigg)^{-3/5}
\bigg(\frac{l_{\rm out}}{10^{24}\,\text{cm}}\bigg)^{2/5}
\bigg(\frac{\bar n_e}{\SI{2.2e-7}{cm^{-3}}}\bigg)^{-6/5}
\bigg(\frac{\nu}{\text{GHz}}\bigg)^{6/5}\;.
\label{eq:rdiff-IGM}
\end{align}
This is comparable to the Fresnel radius for a scintillation screen
\SI{0.25}{Gpc} away, $r_F \simeq \SI{6.3e13}{cm}$. We conclude that
integralactic scintillation is at the borderline between the weak
and strong scintillation regimes. 

\begin{figure}
\centering
\includegraphics[width=0.6\textwidth]{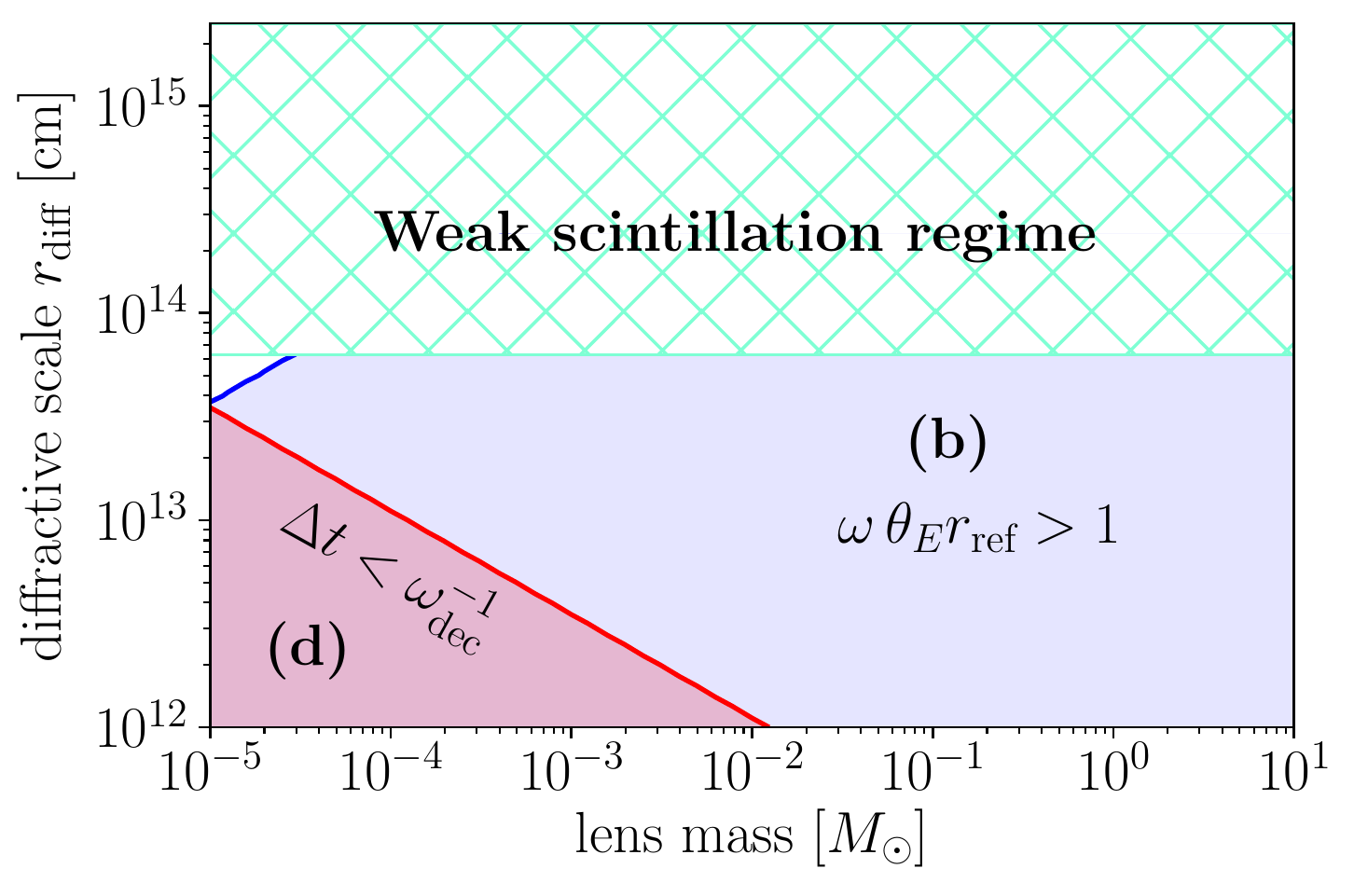}
  \caption{Impact of intergalactic scintillation on FRB lensing by
    compact objects, shown here as a function of the lens mass $M$ and the
    diffractive scintillation scale $r_\text{diff}$.  We assume an FRB at a
    comoving distance of \SI{1}{Gpc}, and a lens at \SI{0.5}{Gpc}.  The
    scintillation screen is placed between the lens and the observer at
    \SI{0.25}{Gpc}.  The angular separation between the lens and the (unlensed)
    source, normalized to the Einstein angle, is $y=\beta/\theta_E=0.5$. The
    frequency of the radio signal is taken to be $\SI{1}{GHz}$.
}
  \label{fig-igm}
\end{figure}

The observability of the lensing signal in the presence of IGM scintillation
is illustrated in \cref{fig-igm} as a function of
$r_\text{diff}$ and of the lens mass. We see that for
realistic parameters and most of the interesting values of the lens
mass the signal will be in regime (b). In other words, the two
lensed images will be distorted incoherently, but the decorrelation
bandwidth of the scintillation will be large enough to accommodate
multiple lensing fringes.  
 
The above estimate is based on the assumption that there are no baryon
overdensities --- such as an interjacent galaxy or galaxy cluster ---
along the line of sight.  As $r_\text{diff}$ scales with $\bar n_e^{-6/5}$,
such an overdensity can greatly increase the strength of the
scintillation. While this might offer interesting
opportunities for studies of the IGM using scintillation~\cite{Ferrara:2001ib,
Pallottini:2013rja, Koay:2014gsa}, it is undesirable for searches of 
FRB diffractive lensing. 
Fortunately, for sources at small redshifts $Z_S< 1$, where most observed
FRBs are expected to occur, the probability that the line of sight
crosses a major baryon overdensity is low. Namely, the probability of
crossing a galaxy cluster is less than $20\%$ and about
$3\%$ for galaxies~\cite{Macquart:2013nba}. 
We will therefore assume the absence of
overdensities and use $r_{\rm diff} \sim \SI{4.5e13}{cm}$ for the IGM.
In a real sample of FRBs, one could use the known locations of galaxies
and galaxy clusters along the various lines of sight to reject those FRBs
which are likely to be affected by baryon overdensities.
Note that this restriction implies that the lens itself cannot lie
inside a galaxy or a cluster. This is not a problem for dark matter,
most of which is distributed in diffuse halos, filaments and voids. On
the other hand, it essentially precludes searching for
diffractive lensing by compact objects of astrophysical origin, such
as brown dwarfs~\cite{Eichler:2017eid}.\footnote{The latter search
  might still be
possible at higher frequencies $\nu\gtrsim 5\,\text{GHz}$ where the
scintillation effects are suppressed and the requirement to avoid
baryon overdensities may be relaxed.} 

\subsection{Scintillation summary}
%--------------------------------------------------

When studying realistic lensed FRB signals, we should in principle consider the
effect of intergalactic scintillation, of interstellar scintillation in the
FRB's host galaxy, and of interstellar scintillation in the Milky Way.  
From the discussion of the previous subsection, we expect the effect
of the IGM to be typically less important than that of the ISM. We have
verified this using a sample of simulations where we combined IGM
scintillation screens with ISM screens (see \cref{sec:simulation-and-results}
for the description of the simulation code \cite{github}). However,
the numerical cost of the simulations grows rapidly with the number of
screens, so to reduce the computation time, we {\em neglect} IGM
scintillation in the rest of our analysis. 

The properties of the FRB host galaxies remain largely unknown, despite
localization of several burst sources
\cite{Chatterjee:2017dqg, Tendulkar:2017vuq, Bannister565, Ravi:2019alc}.
In the absence of any detailed information, we assume that the
properties of the interstellar medium in the host galaxy are similar
to that in the Milky Way. Then, barring a chance alignment of the line
of sight with the galactic disk of the host, one can assume the
scintillation screen of the host galaxy to have similar a effect on the
lensing signal as the ISM screen of the Milky Way.\footnote{Strong
  temporal broadening of the signal has been observed for some FRBs and is
  consistent with scattering on dense plasma inhomogeneities.  These are likely
to occur in the close vicinity of the 
source~\cite{Katz:2015mpa,Xu:2016vjj}, in which case they do
not significantly affect the detectability of lensing.} 
Therefore, for our
exploratory study of the detectability of FRB lensing we choose to
consider only the ISM screen of the Milky Way, which allows us to
further reduce the required computing resources. It is worth
emphasizing, however, that an actual future search for lensing signal
in the data should include modeling of scintillation both in the
host galaxy and in the IGM.

%==========================================================================
\section{Simulations and Results}
\label{sec:simulation-and-results}
%==========================================================================

%  !TEX root =  draft.tex
%--------------------------------------------------------------------------
\subsection{Formalism}
\label{sec:sim-formalism}
%--------------------------------------------------------------------------

To better understand the effect of scintillation on the spectra of astronomical
radio sources in general and FRBs in particular, we have developed a simulation
code to describe both lensing and scintillation.
It is capable of evolving the signal from a
point-like or extended radio source through an arbitrary stack of lensing and
scintillation screens, and onwards to an observer.  Each screen corresponds to a
two-dimensional plane, oriented perpendicular to the line of sight. It associates
each point $\vec{x} = (x^1, x^2)$ on the plane with a characteristic phase factor
$\varphi_j(\vec{x})$ that is imparted to photons passing through that point on the
screen. Here, the index $j$ numbers the successive screens.
The photon field
$A_j(\vec{x})$ on the $j$-th screen is obtained from the field $A_{j-1}(\vec{x})$ on
the previous screen according to the Fresnel integral, 
\begin{align}
  A_j(\vec{x}) = \tfrac{k \, e^{i k (z_j - z_{j-1})}}{2\pi i (z_j - z_{j-1})}
                 \int\!d^2x' \, A_{j-1}(\vec{x}') \exp\Big(
                   \tfrac{i k |\vec{x} - \vec{x}'|^2}{2 (z_j-z_{j-1})} + i \varphi_j(\vec{x})
                 \Big) \,
  \label{eq:fresnel-2d}
\end{align}
(in analogy to \cref{eq:signalO,eq:phi}).
Here, $z_j$ is the comoving line-of-sight distance of the source and the $j$-th screen,
and the wave number $k$ is given by the dispersion relation in a
plasma, \cref{eq:dispersion}.  In evaluating the plasma frequency $\omega_p$, we assume
a mean ISM plasma density of $n_e = \SI{0.03}{cm^{-3}}$.  The plasma density in the IGM is
several orders of magnitude smaller, but as we neglect scintillation in the IGM
following the discussion in \cref{sec:ism-igm}, its precise value does not affect
our analysis.  Without IGM scintillation, there is no multi-path propagation through the IGM,
so all photons of a given frequency are dispersed by the same amount.
The screen $A_0(\vec{x})$ corresponds to the source; for a point-like source, it is
simply a $\delta$-function centered at the coordinates $\vec{x}_0$ of the source,
$A_0(\vec{x}) = \delta^{(2)}(\vec{x} - \vec{x}_0)$.  The last ($n$-th) screen in the stack is
placed at the location of the observer, so that the photon field $A_n(\vec{x})$
corresponds to the observed image.  We will not consider the spatial extent
of the image here and just evaluate $A_n(0)$ for simplicity.

As in \cref{sec:scint-primer}, the first term in the exponential factor under
the integral in~\cref{eq:fresnel-2d} describes the additional geometric phase
that photons experience when traveling from $\vec{x}'$ to $\vec{x}$ (in the
limit that $|\vec{x}| \ll z_j - z_{j-1}$).  For lensing screens, the phase
factor $\varphi_j(\vec{x})$ describes the non-geometric part of the Shapiro
time delay
\begin{align}
  \varphi_j^\text{lens}(\vec{x}) = -\frac{(z_{j+1} - z_j)(z_{j+1} - z_{j-1})}{z_j - z_{j-1}}
                                   \, \psi(\vec{x}) \,,
  \label{eq:phi-lens}
\end{align}
where $\psi(\vec{x}) = \theta_E^2 \log\big[ |\vec{x}| / (z_{j+1} - z_{j-1}) \big]$ is again
the lensing potential for a point-like lens, and $\theta_E$ is the Einstein
angle defined in Eq.~\eqref{eq:EinsteinAngle}.

For scintillation screens, $\varphi_j(\vec{x}) =
\varphi_j^\text{scint}(\vec{x})$ describes the variation in the dispersion
relation due to random fluctuations of the plasma density in the scintillating
medium.  As discussed in \cref{sec:ism-igm}, it is the modeling of these
fluctuations that is the main source of uncertainty in our calculations. We
assume Kolmogorov turbulence~\cite{Kolmogorov:1941a, Kolmogorov:1941c,
2004ARA&A..42..211E}, i.e.\ we assume that the fluctuation mode
with wave number $\vec\kappa$ on the screen has an amplitude
$|\tilde{f}(\vec\kappa)|$ drawn from a Gaussian distribution of 
width\footnote{In general the statistics of the scintillation phase may be
  non-Gaussian.  A study of the corresponding effects is, however, beyond the scope
of this paper.}
\begin{align}
  \sigma^\text{scint}(\vec\kappa)
    = \bigg(\frac{\vec\kappa^2}{\kappa_\text{diff}^2}\bigg)^{11/6} \,.
  \label{eq:kolmogorov-spectrum}
\end{align}
Here, $\kappa_\text{diff} \equiv 2 \pi / r_\text{diff}$ is an arbitrarily chosen
reference scale.  The phase of
$\tilde{f}(\vec\kappa)$ is also a random number, uniformly distributed between
$0$ and $2\pi$.  We emphasize that $\vec\kappa$ describes density fluctuations
on a scintillation screen and should not be confused with the wave number $k$
of the photon field.  The scintillation phase $\varphi_j^\text{scint}(\vec{x})$
is then proportional to the Fourier transform of $\tilde{f}(\vec\kappa)$:
\begin{align}
  \varphi_j^\text{scint}(\vec{x}) \propto
    \int \! d^2\kappa \, e^{i \vec\kappa \vec{x}}
    \tilde{f}(\vec\kappa) \,.
  \label{eq:scint-ft}
\end{align}
The normalization of $\varphi_j^\text{scint}(\vec{x})$ is chosen such that $\mathcal{O}(1)$
variations in the phase occur over distance scales of order $r_\text{diff}$,
\begin{align}
  \ev{\big[\varphi(\vec{x}) - \varphi(\vec{x} + r_\text{diff} \hat{n})\big]^2} = 1 \,,
  \label{eq:scint-norm}
\end{align}
where $\ev{\cdot}$ denotes averaging over the screen, and $\vec{n}$ is a
unit vector, which we choose to be oriented in the $x^1$ direction.\footnote{This
  choice is convenient, given our discretization of the scintillation screen.
  In principle, we should average over all directions of $\vec{n}$, but doing so
would not significantly affect our results because of the statistical isotropy
of the screen.}

We assume the incoming amplitude of the radiation at frequency $\omega$ to be of 
the form:
\begin{align}
  f_\text{in}(\nu) \propto \exp\bigg(\!\! -\frac{\nu}{\SI{200}{MHz}} \bigg)
  \label{eq:input}
\end{align}
This expression is an attempt to roughly fit an observed FRB spectrum from
ref.~\cite{Masui:2015kmb}.
While we do not claim that our extrapolation to higher frequencies than observed
in ref.~\cite{Masui:2015kmb} is necessarily valid, our further 
analysis will rely mainly on a reconstruction of the transfer function,
largely reducing the uncertainties associated with the 
exact shape of the FRB spectrum (see \cref{sec:data-analysis}).

%--------------------------------------------------------------------------
\subsection{Numerical Implementation}
\label{sec:numerics}
%--------------------------------------------------------------------------

To compute the Fresnel integral in \cref{eq:fresnel-2d} numerically, we
employ several simplifications:
\begin{enumerate}
  \item As the Fresnel integral on a lensing screen is numerically very badly
    behaved, we treat lensing analytically.  In other words, if the $j$-th screen
    is a lensing screen, we evolve the photon field directly from screen $j-1$
    to screen $j+1$, multiplying by the appropriate magnification, see \cref{eq:split}.
    For the lens masses considered in this paper, we always work in the
    geometric optics limit, even though our code (released together with
    this paper~\cite{github}) is also able to handle the wave optics
    regime, and switches to it automatically if
    $\omega (D_L D_S / D_{LS}) \, \theta_E^2 < 10$.

  \item We discretize scintillation screens, i.e.\ we evaluate the
    integrand in \cref{eq:fresnel-2d} on a square grid of points, and then evaluate
    the integral as a
    simple sum.  The size of the screen in both directions is chosen $\gg
    r_\text{ref}$ to make sure the region $|\vec{x}| \lesssim r_\text{ref}$, from
    which most of the signal is received, is well within the simulated region.
    (At distances $\gg r_\text{ref}$ from the line of sight, the geometric
    phase varies much faster than the scintillation phase, so that photons
    emitted from this regions mostly interfere destructively.) The grid spacing
    $dx$ is chosen much smaller than $r_\text{diff}$ to make sure the phase
    fluctuations on the screen are well resolved. Moreover, we require $dx$
    to be small enough to still resolve the fast-varying geometric phase at the
    edges of the screen. (The last requirement is dropped when the grid becomes
    so large that it would not fit into GPU memory any more.)

    To further optimize memory consumption, we do not simulate the whole screen
    at once. Instead, we first simulate the long wavelength fluctuations,
    discretized with a coarser grid spacing. The geometric phase factor in
    the Fresnel integral from \cref{eq:fresnel-2d} is evaluated only once
    for each cell on this coarse grid.    The cell size of the coarse grid is
    chosen $\ll r_F$ to ensure that this is a good approximation.  For
    simulating shorter wavelength fluctuations, we divide the screen into
    smaller patches and simulate the fluctuations on each of them separately,
    but at full resolution $dx$. The patches are always chosen much
    larger than $r_\text{diff}$.
    
  \item For most of our results, we reduce the scintillation screen to
    one dimension. In other words, we assume all photons to travel in the
    $x^1$--$z$ plane.  This greatly reduces memory consumption and computational
    effort, while all the essential features of scintillation are preserved,
    as evidenced by the good agreement with the qualitative analytic
    estimates from \cref{sec:separation}.
\end{enumerate}
Our code~\cite{github} is written in Python~3, making heavy use of the \texttt{CuPy} 
library~\cite{cupy_learningsys2017} to run it on CUDA-enabled GPUs and thus optimally
benefit from the vectorizability and scalability of our approach.
In order to vectorize the evaluation of the magnification function for gravitational
lensing in the wave optics regime, we had to implement vectorized versions of
certain special functions, notably the gamma function $\Gamma(z)$ and the Laguerre
polynomials $L_n(z)$, for the case of complex $z$ and $n$.  In implementing
the Gamma function, we follow Lanczos' approximation \cite{Lanczos:1964,Press:NumRecip},
while our implementation of the Laguerre polynomials is based on the implementation
in the \texttt{mpmath} package~\cite{mpmath}, see also~\cite{johansson:2016}.

%--------------------------------------------------------------------------
\subsection{Data Analysis}
\label{sec:data-analysis}
%--------------------------------------------------------------------------

Let us now outline the procedure we use to extract the lensing signal from our
simulated data.  Our objective is to identify the periodic modulation of the
frequency spectrum due to lensing on top of the stochastic background from
scintillation. We would like to work in particular with the Fourier
transform
\begin{align}
  |\mathcal{T}(\tau)| \equiv \big| \operatorname{FFT}\big( |T(\nu)|^2 \big) \big|
  \label{eq:T-FFT}
\end{align}
of the transfer
function, \cref{eq:transfer-function}, where the lensing signal materializes as a peak at
$\tau = \Delta t$.  Of course, we do not have direct access to the transfer function
and its Fourier transform as we do not know the initial spectrum at the source.
Moreover, observing the lensing peak is complicated by the fact that it is
typically located on top of a steeply falling (or rising) background.

The first step in our analysis chain is a fit of the logarithm of the
``observed'' radiation intensity, $\log I_\text{obs}(\nu) \equiv \log
|f_\text{obs}(\nu)|^2$ with a high-order polynomial (we use $d = 15$).  To
ensure the fit is smooth, we use polynomial lasso regression with
regularization parameter $0.05$~\cite{Santosa:1986, Tibshirani:1996}.  We then
subtract the fit function from $\log I_\text{obs}(\nu)$ to obtain an estimate
$\log |\hat{T}(\nu)|^2$ of the log of the transfer function
$|T(\nu)|^2$,~\cref{eq:transfer-function}.  Besides lasso regression with a
15th order polynomial, we tried several alternative fitting functions and
regularization methods (for instance ridge regression). We found them to all
perform roughly comparably. The minor effects that various fitting strategies
might induce on the reconstructed spectrum, can be studied by the reader
independently using our public code~\cite{github}.  

\begin{figure}
  \centering
  \includegraphics[width = 1.05\textwidth]{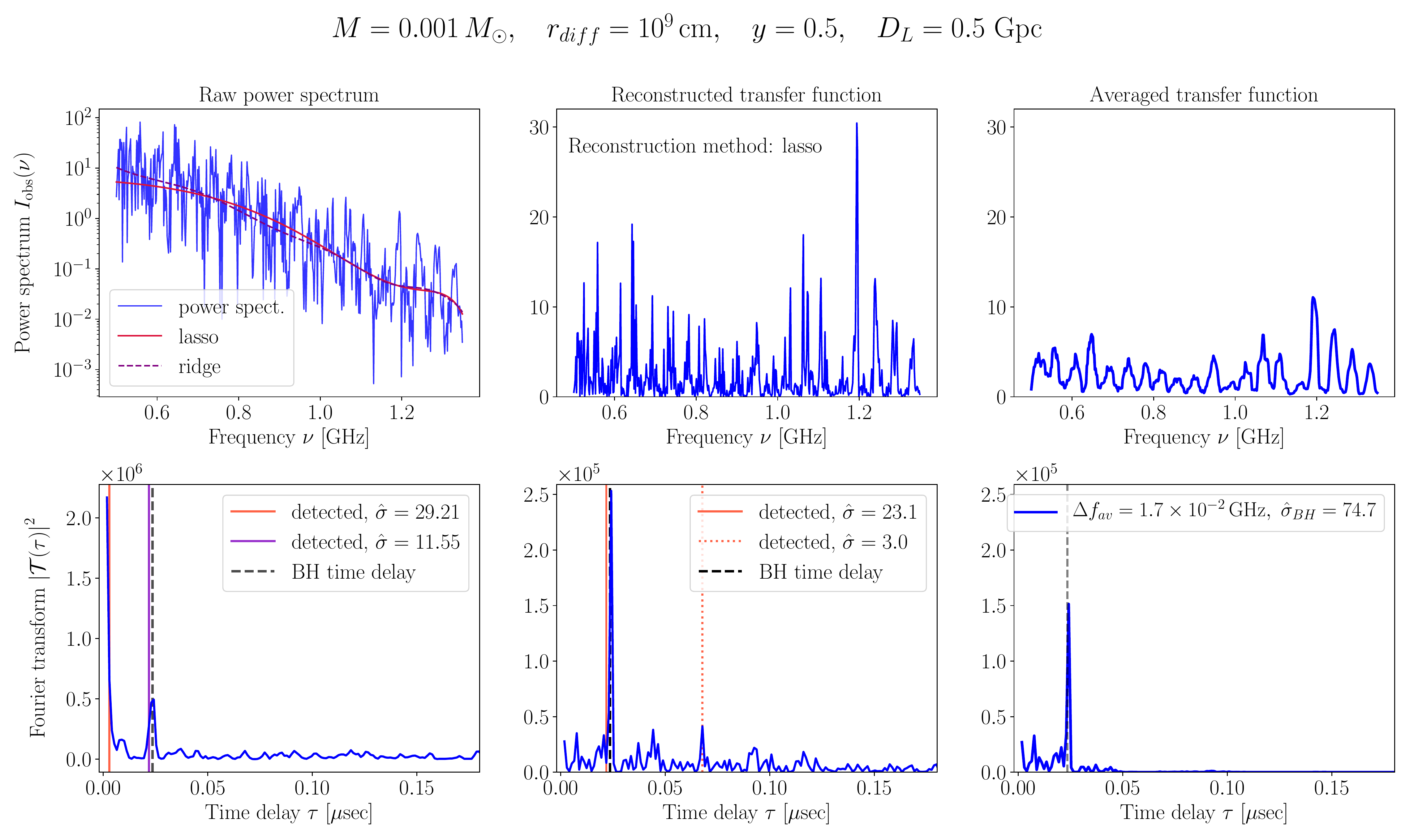}
  \caption{The analysis of a single, diffractively lensed FRB spectrum
    following the steps described in the text. The upper left plot shows the
    radiation power spectrum, and the lower left plot shows the Fourier
    transform thereof, where the lensing peak is clearly visible at the
    expected time delay (vertical black line). However, we also notice a strong
    enhancement of the power at small $\tau$, which in some cases spoils the
    detectability of the lensing peak.
    Similarly, the middle column shows the reconstructed
    transfer function and its Fourier transform, and the column on the right
    shows the same after averaging procedure described in the text.
    The enhancement at small $\tau$ has disappeared and only the lensing peak
    remains.  We
    have assumed a source at a comoving distance $D_S = \SI{1}{Gpc}$ and a lens
    at $D_L = \SI{0.5}{Gpc}$, $y = 0.5$.  An ISM scintillation screen is placed
    at $D_\text{ScO} = \SI{1}{kpc}$, and $r_\text{diff} = \SI{e9}{cm}$ at the
    spectral midpoint.
  }
  \label{fig:one_event_analysis} 
\end{figure}

At the second stage of our analysis, we look for peaks in $|\mathcal{T}(\tau)|$,
defined in \cref{eq:T-FFT}.  For a fixed
peak time delay $\tau_0$ and peak width $w_\tau$, we fit a polynomial
$P(\tau_0, w_\tau; \tau)$ to the Fourier power spectrum
(in linear space), excluding the points in the
interval $[\tau_0 - w_\tau/2, \tau_0 + w_\tau/2)$.  The fitting is performed by
polynomial ridge regression, with regularization parameter $10^{-4}$.  As
previously, fitting with other parameters yields comparable results and the
small differences can be traced by using alternative methods in our
code~\cite{github}. 
We then define a test statistic $\hat\sigma$ as the difference between
$|\mathcal{T}(\tau)|^2$ at the peak and the polynomial at this time delay,
normalized by the width of the peak and by the standard deviation of
all samples outside the peak region:
\begin{align}
  \hat\sigma(\tau_0, w_\tau) = \frac{N_{\rm out~peak}\sum_\text{$j$ in peak}
                                   \big( |\mathcal{T}(\tau)|^2 - P(\tau_0, w_\tau; \tau) \big)}
                                    {N_{\rm in~peak} \sum_\text{$j$ outside peak}
                                   \big| |\mathcal{T}(\tau)|^2 - P(\tau_0, w_\tau; \tau) \big|} \,.
  \label{eq:test-statistic}
\end{align}
Here the sum in the numerator runs over the $N$ time delay samples inside the
peak region, while the sum in the denominator runs over all samples outside the
peak region.  We repeat this procedure for each time delay $\tau_0$ and each
width $w_\tau$, and we consider as a potentially interesting peak any local
maximum in the two-dimensional manifold defined by $\hat\sigma(\tau_0, w_\tau)$.
We quote a non-zero test statistic for the lensing signal if a peak is found
located within 15\% from the ``real'' $\Delta t$ of the lens. Otherwise, we set
$\hat\sigma = 0$. Note that the
lensing peak is typically narrow, so the highest significance is always found
for the minimum value of $w_\tau$, which corresponds to a peak spanning two
time delay bins. The procedure is illustrated in \cref{fig:one_event_analysis}.

In the right-hand column of that figure, we also show how the lensing peak
in $|\mathcal{T}(\tau)|^2$ is affected if we average $I_\text{obs}(\nu)$
over frequency intervals $\Delta \nu_\text{av} = \SI{17}{MHz}$
before taking the Fourier transform.  We see that high-$\tau$
fluctuations are suppressed, but also the height of the lensing peak is
somewhat reduced.  Averaging is therefore a useful tool only when high-$\tau$
fluctuations are worrisome.  In our subsequent analysis, we will not use
averaging.

\begin{figure}
  \centering
  \includegraphics[width = 0.75\textwidth]{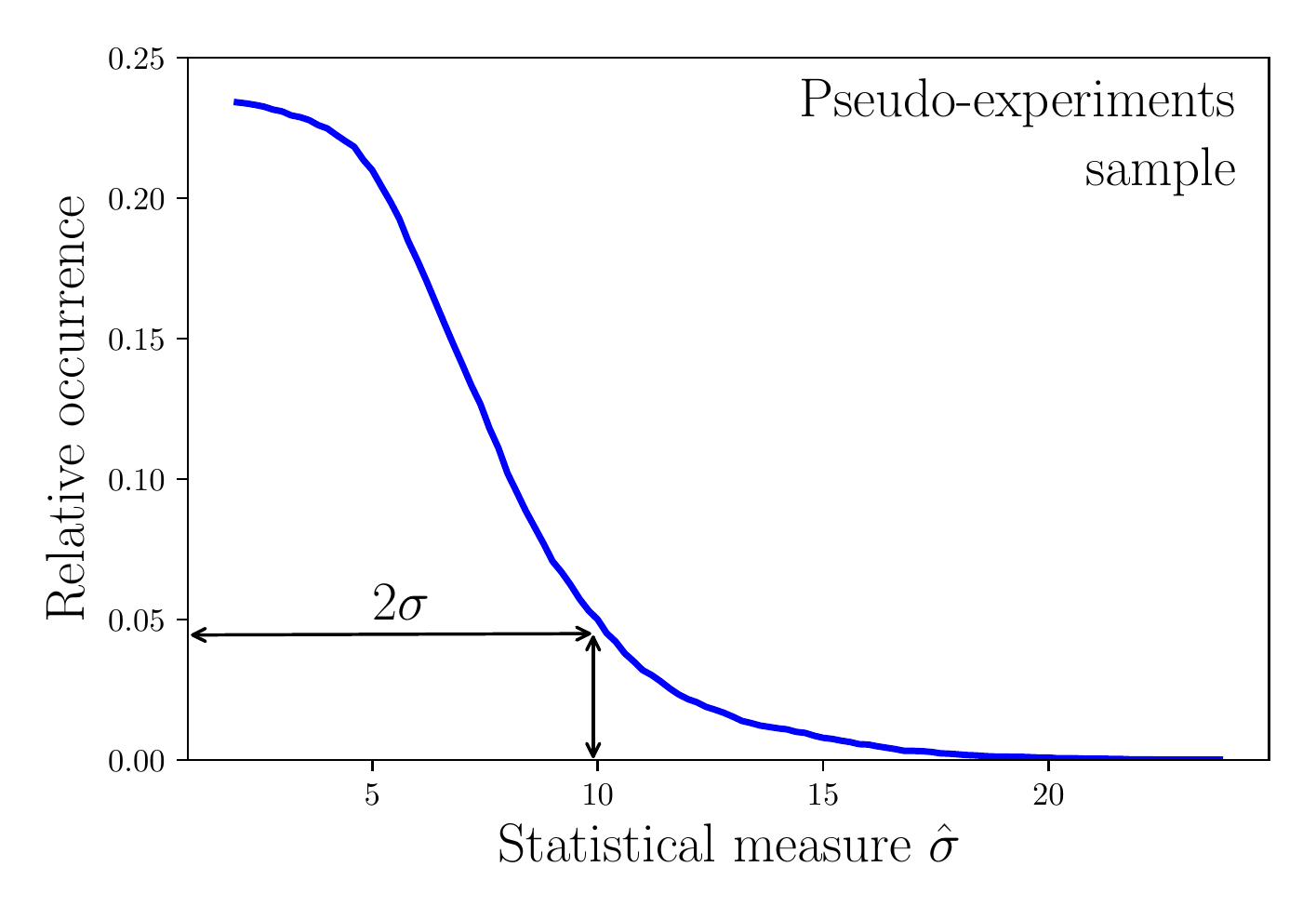}
  \caption{Relative occurrence of statistical measures $\hat \sigma$ in the 
    FFT-squared of the power spectrum in the pseudo-experiments sample. 
    The distribution can be reasonably approximated by Gaussian. We calibrate 
    our statistical significance based on this curve.  }
  \label{fig:calibration}
\end{figure}

To translate the test statistic $\hat\sigma$ into a statistical significance,
which can then be interpreted as a quantile of the normal distribution,
we have simulated 10\,000 pseudo-experiments. Each of them corresponds to
an unlensed FRB distorted
by a scintillation screen with $r_\text{diff} = \SI{e10}{cm}$ at \SI{1}{kpc}.
In each pseudo-experiment, we look for a spurious peak at the $\Delta t$
that would correspond to lensing by a \SI{e-3}{M_\odot} PBH
located at $D_L = \SI{0.5}{Gpc}$, $y=0.5$.  We plot the distribution of
the $\hat\sigma$ values of the spurious peaks in \cref{fig:calibration}.
The quantiles of this
distribution correspond to the real statistical significance $\sigma$, which
we will use below to set limits on the PBH abundance.

One comment is in order regarding this calibration procedure. The
distribution of $\hat\sigma$ values can depend on the diffractive scale
$r_\text{diff}$ of the scintillation screen. Nonetheless, as generating a large
number of pseudo-experiments is computationally expensive, we use the curve in
\cref{fig:calibration} for all values of $r_\text{diff}$.  For $r_\text{diff}
> \SI{e10}{cm}$, applying calibration based on \cref{fig:calibration} is
too conservative, while for smaller $r_\text{diff}$, it may somewhat
overestimate the sensitivity.  However, we are going to find that for
$r_\text{diff} \sim \SI{e9}{cm}$, the sensitivity is weak anyway.

%--------------------------------------------------------------------------
\subsection{Results}
\label{sec:results}
%--------------------------------------------------------------------------

\begin{figure}
  \centering
  \includegraphics[width = .99\textwidth]{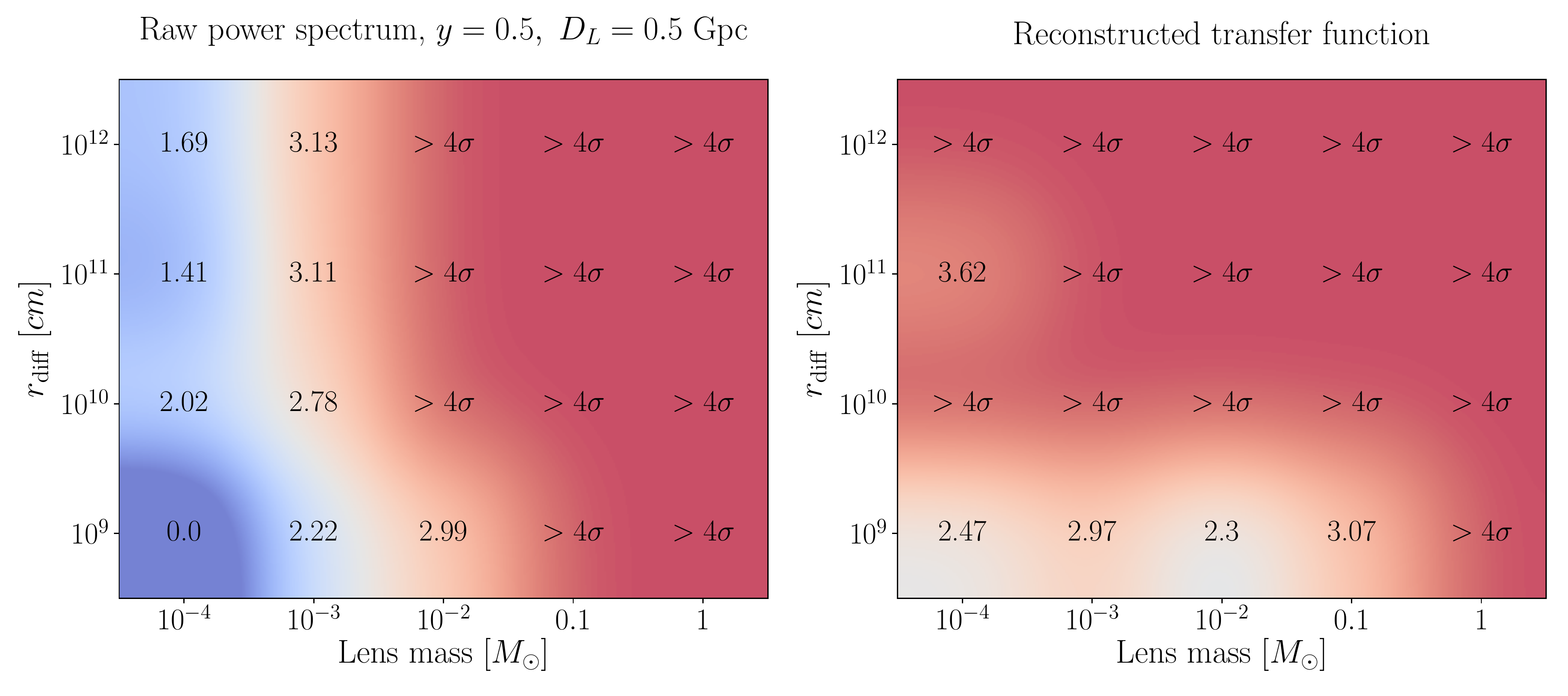}
  \caption{The local statistical significance of the diffractive lensing signal
    expected in a typical FRB spectrum as a function of the lens mass $M$ and
    the diffractive scintillation scale $r_\text{diff}$ of the interstellar
    medium.  We assume an FRB at a comoving distance of \SI{1}{Gpc} and a lens
    at $D_L = \SI{0.5}{Gpc}$, $y = 0.5$. An ISM scintillation screen is placed
    at $D_\text{ScO} = \SI{1}{kpc}$, and the quoted $r_\text{diff}$ values are
    at the spectral midpoint.  Numbers in each grid cell indicate
    the local significance of the scintillation peak in our simulations,
    calculated as described in \cref{sec:data-analysis}.  The left-hand panel
    is based directly on the simulated radiation intensity spectrum, while in
    the one on the right, we have divided out the smooth component of the
    spectrum using lasso regression.  A clear benefit of
    this step is visible, especially at low lens masses.
    }
  \label{fig:mrscan}
\end{figure}

To anticipate the sensitivity of FRB lensing to compact dark matter
objects, we now investigate the significance of the lensing signal as a
function of the model parameters, in particular the lens mass $M$, its position
$D_L$ along the line of sight and angular distance $\beta$ from the line of
sight, as well as the diffractive scintillation scale $r_\text{diff}$ in the
Milky Way.  For simplicity, we will keep the distance of the source fixed at
\SI{1}{Gpc} (comoving), we will neglect intergalactic scintillation as well as
scintillation in the FRB's host galaxy, and we will assume the Milky Way's
scintillation screen to be located at a distance of \SI{1}{kpc}.

As a first result, we show in \cref{fig:mrscan} the local statistical
significance of the lensing signal as a function of $M$ and $r_\text{diff}$,
keeping $D_S = \SI{1}{Gpc}$, $D_L = \SI{0.5}{Gpc}$, and $y \equiv \beta/\theta_E = 0.5$
fixed.
Note that in this scan we choose the instrumental frequency band and
frequency resolution as needed to well resolve the lensing peak.  This assumption
may be unrealistic at very small lens masses, which require very large
frequency bands, and at very large lens masses, which require very high
frequency resolution. 
Therefore, at high and low lens
masses \cref{fig:mrscan} should be thought of as proof of principle rather than
as a realistic expectation.  \Cref{fig:mrscan} confirms our expectations from
\cref{fig:scint-regions}: the sensitivity is good except at the 
smallest $r_\text{diff}$ values.  In the left panel of \cref{fig:mrscan}, we
observe a loss of sensitivity at low masses.  There, the lensing
peak in $|\mathcal{T}(\tau)|^2$ falls in the low-$\tau$ region and is buried
underneath the Fourier transform of the smooth component of the intensity
spectrum.  As expected, reconstruction of the transfer function by dividing out
the smooth component significantly increases the sensitivity, as illustrated
in the right-hand panel.

In order to proceed further, we focus specifically on a future
SKA-type observatory\footnote{https://www.skatelescope.org}. 
This is expected to have a wide frequency band
extending above a few GHz. Going to higher frequencies
is beneficial for detecting
diffractive 
lensing by dark matter as it makes scintillation less important.
Of course, our analysis can be repeated for radiotelescopes
operating at lower frequencies, in particular, for 
CHIME\footnote{https://chime-experiment.ca} and
ASKAP.\footnote{https://www.atnf.csiro.au/projects/askap/index.html. } 

As our fiducial parameters we will assume a frequency resolution no better than
\SI{1}{MHz} and a band width no larger than \SI{10}{GHz}.  This is based on the
frequency range of the SKA-MID array, and on the expected resolution
of the instrument 
\cite{SKA:2015, Bonaldi:aa}.  With these assumptions, we simulate the
lensing signal as a function of the lens mass, the lens distance $D_L$
from the observer 
and its angular distance $y$ from the line of
sight. Once again, we assume $D_S = \SI{1}{Gpc}$, and we introduce an
interstellar scintillation screen at \SI{1}{kpc}, with $r_\text{diff} =
\SI{e10}{cm}$. 
For each combination of $M$, $D_L$ and $y$, we then estimate
the expected local significance of the signal, as described in
\cref{sec:data-analysis}. 
Note that we do not run simulations for 
lens masses above \SI{0.1}{M_{\odot}} because detecting a lensing
signal in that case would require 
a frequency resolution better than \SI{1}{MHz}.  Similarly, we do not
consider lens masses below \SI{e-4}{M_\odot}, where the spacing of
lensing fringes in the spectrum becomes comparable to the maximum bandwidth.

To obtain a bound on the abundance
of point-like lenses such as PBHs, we largely follow the statistical
procedure described in~\cite{Katz:2018zrn}.
We define the likelihood function for a single FRB source at redshift $Z_S$,
\begin{align}
  L_\text{FRB}(M, \rho_\text{PBH}, Z_S)
    = L_0(0) +  \int\!d^3x \, (1+Z_L)^3 \, \frac{\rho_\text{PBH}}{M}
                              \big[  L_0 ( \vec{\mu} ) - L_0 (0) \big] \,.
  \label{eq:L-FRB}
\end{align}
Here, $L_0(\vec{\mu})$ is the likelihood of the (simulated) data -- assumed not
to contain a lensing signal -- when compared to a model that does contain a
lens with parameters $\vec{\mu} \equiv (M, Z_L, y)$.  We set $L_0(\vec\mu) = 1 -
\operatorname{erf}(\sigma/\sqrt{2})$, where $\sigma$ is the significance of the
lensing signal discussed in \cref{sec:data-analysis} above.  This immediately
implies that $L_0(0)$, the likelihood of the data when compared to the
theoretical prediction in the absence of a lens, equals unity.
$L_\text{FRB}(M, \rho_\text{PBH}, Z_S)$ then measures the likelihood of data
which does not contain a lensing signal when compared to the prediction
assuming a population of lenses of mass $M$ and mass density $\rho_\text{PBH}$.
The factor $d^3x \, (1+Z_L)^3 \, \rho_\text{PBH} / M$ measures the probability
of finding a lens in a small volume element $d^3x$ at redshift $Z_L$.
The integral in \cref{eq:L-FRB} runs from the source to the observer
in the longitudinal ($z$) direction, and out to infinity in the transverse
direction. (In practice, we cut it off at $y = |\vec{x}_\perp|/(D_L \theta_E) = 5$,
as there will be no sensitivity to lenses that are further away from the line of sight.)
Note that $\vec{x}$ here refers to a three-dimensional physical (not co-moving)
coordinate.  In writing down \cref{eq:L-FRB}, we have assumed that the probability
for a single source to be lensed by two lenses simultaneously is negligible.

To compute $L_\text{FRB}(M, \rho_\text{PBH}, Z_S)$ for a given lens mass $M$,
lens density $\rho_\text{PBH}$, and source redshift $Z_S$, we evaluate
$\sigma$ as function of the lens coordinates.  We then evaluate
the integral over $d^3x$ on the resulting grid.

Of course, observing just a single FRB is not sufficient to set a meaningful
limit on compact dark matter.  Obvserving a large number $N$ of them, however,
will greatly boost the sensitivity of the method.  We estimate the expected
95\% confidence level exclusion limit on $\rho_\text{PBH}$ for fixed $M$ by
solving the equation
\begin{align}
  -2 \log \Bigg( \prod_{i=1}^N \frac{L^i_\text{FRB}(M, \rho_{PBH}, Z_S^i) } 
                                    {L^{i}_\text{FRB}(0, 0, Z_S^i)} \Bigg)
  = 5.99 \,, 
\end{align}
Note that $L^{i}_\text{FRB}(0, 0, Z_S^i) = 1$.
For simplicity, we will in the following assume that all sources are at the
same redshift, and that also the properties of the interjacent scintillation
screens are the same.  When the probability for lensing a single FRB is small
($L_\text{FRB}(M, \rho_\text{PBH}, Z_S^i) \approx 1$), the reach
for the PBH density $\rho_\text{PBH}$ is then inversely proportional to the number
of sources $N$.  This is because increasing the number of sources linearly
decreases the effective optical depth.

\begin{figure}
  \centering
  \includegraphics[width=1.05\textwidth]{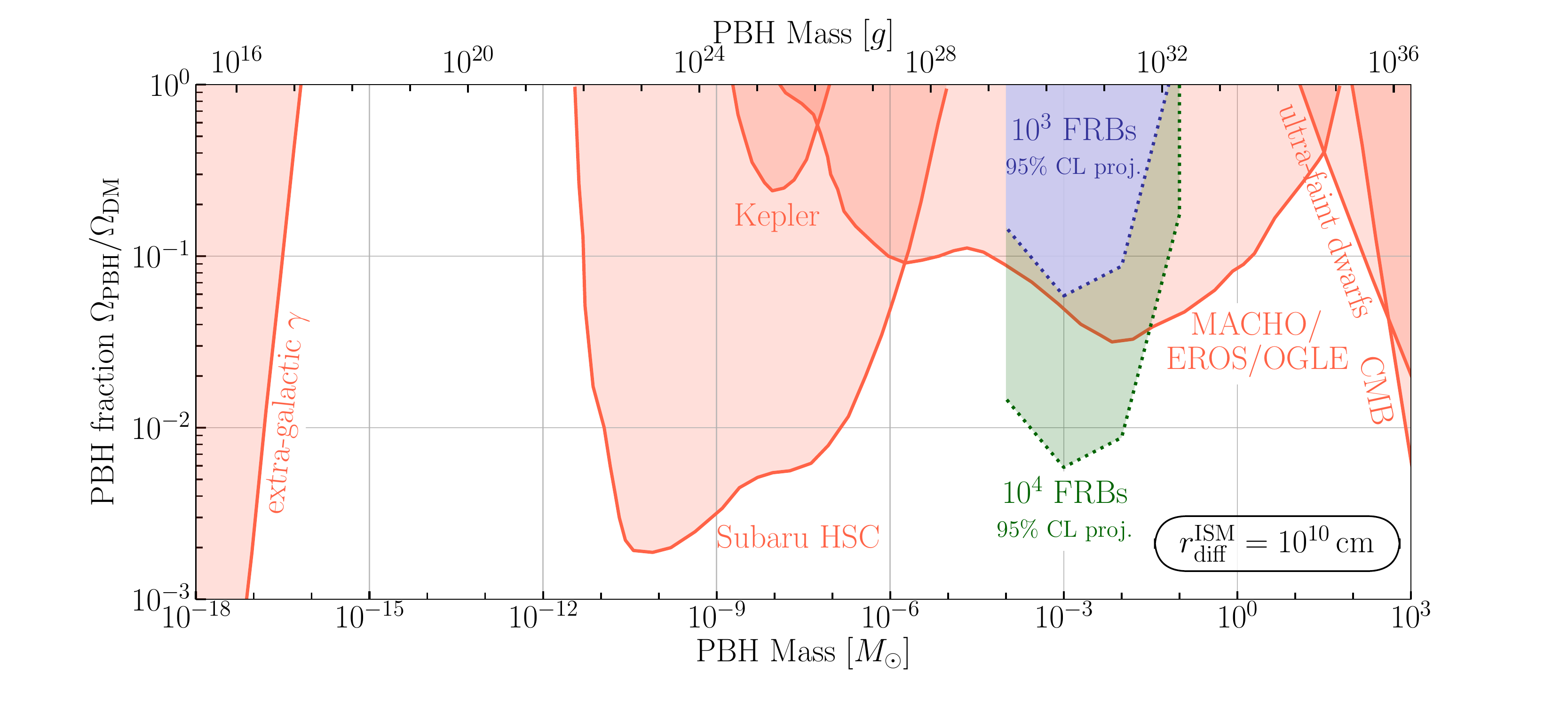}
  \caption{Expected limits on primordial black holes
    from diffractive lensing of FRBs. We assume an interstellar scintillation screen
    at a distance of \SI{1}{kpc}, with diffractive scale $r_\text{diff}^\text{ISM} =
    \SI{e10}{cm}$ at \SI{1}{GHz}, and negligible intergalactic scintillation.
    We compare our projections (in blue) to
    microlensing constraints from Subaru
    HSC~\cite{Niikura:2017zjd},\protect\footnotemark\
    Kepler~\cite{Griest:2013esa}, MACHO~\cite{Allsman:2000kg},
    EROS~\cite{Tisserand:2006zx}, and OGLE~\cite{Wyrzykowski:2011tr}, to CMB
    constraints due to accretion onto PBHs~\cite{Ali-Haimoud:2016mbv}, to
    limits based on the dynamics of ultra-faint dwarf
    galaxies~\cite{Brandt:2016aco}, and to bounds on the contribution of
    Hawking radiation from PBHs to the extra-galactic gamma
    background~\cite{Carr:2009jm}.  We emphasize that CMB and gamma-ray
    background constraints are based on specific properties of PBHs, and that
    the other existing constraints probe the \emph{local} distribution of dark
    matter in the Milky Way and its immediate neighborhood.  By contrast,
    diffractive lensing of FRBs offers a model-independent probe of MACHOs
    at cosmological distances.
  }
  \label{fig:money}
\end{figure}

The resulting bound on the PBH parameter space is shown in~\cref{fig:money} and
compared to other constraints on primordial black holes.
We see that, for a realistic number of FRB observations in SKA (a few tens
per day \cite{Macquart:2015uea}), our method can be
expected to yield highly competitive constraints.  Even more importantly, the
expected limits are complementary to those from microlensing searches
because FRBs probe the distribution of dark matter over cosmological distance scales,
while microlensing is sensitive to dark matter in the Milky Way and its immediate
vicinity.  This makes diffractive lensing of FRBs a particularly interesting probe
for compact dark matter mini-halos, which may be inaccessible to microlensing searches
due to their non-negligible spatial extent \cite{Dror:2019twh}.  Indeed, the
condition $r_L < r_E$ on the lens size $r_L$, under which the lens can be treated
as point-like, is more readily satisfied at cosmological distances.
Besides, mini-halos may exist in regions of low baryon density, but may have
suffered tidal disruption in the Milky Way.  The
sensitivity of diffractive lensing ends at PBH masses below $\SI{e-4}{M_\odot}$
because at such low masses the lensing peaks in the frequency spectrum are too
far apart for a significant number of them to be contained in a realistic
instrumental frequency band.  
At high PBH mass, the limiting factor is the
instrumental frequency resolution.

\footnotetext{See also the recent discussion of finite source size effects on the Subaru~HSC
      constraint~\cite{Smyth:2019whb, Montero-Camacho:2019jte}.}

In general, it should be mentioned, that in vast majority of the models the
mass distribution of the dark matter can be relatively broad. While we are
assuming here a very narrow distribution, generalization to realistic mass
distributions is desirable~\cite{Green:2016xgy, Kuhnel:2017pwq, Carr:2017jsz,
Bellomo:2017zsr, Laha:2018zav,Poulter:2019ooo}.

Finally, we briefly mention other proposals for probing PBHs in the mass range
between \SI{e-4}{M_\odot} and \SI{1}{M_\odot}.  A novel method to probe
compact objects in the Milky Way is pulsar timing~\cite{Dror:2019twh}.
An alternative, indirect, approach involves primordial gravitational waves
produced by adiabatic density fluctuations which also give rise to PBH formation
\cite{Saito:2008jc, Garcia-Bellido:2017aan}.  The method is thus specific to PBH production via
enhanced adiabatic Gaussian density fluctuations.\footnote{Recently, this
  method has been applied to NANOGrav data~\cite{Chen:2019xse}.  However, the
resulting limits are exponentially sensitive to theoretical uncertainties.}
At somewhat larger PBH masses, one can look for gravitational waves from PBH
mergers \cite{Kavanagh:2018ggo, Chen:2019irf, Authors:2019qbw}.  The expected
gravitational wave signal is, however, subject to large uncertainties in
the evolution of PBH populations over cosmological history
\cite{Nakamura:1997sm, Ioka:1998nz, Ali-Haimoud:2017rtz, Kavanagh:2018ggo, Raidal:2018bbj}.
In the region of lighter masses, PBHs can lead to the destruction of
neutron stars~\cite{Capela:2013yf} and white dwarfs~\cite{Graham:2015apa}.
Exploting these mechanisms to set limits is, however, complicated by
uncertainties in the dark matter concentration in globular clusters~\cite{Conroy:2010bs,
Ibata:2012eq} and in the treatment of energy deposition by PBHs crossing
stars~\cite{Montero_Camacho_2019, Katz:2019rgf}.

%==========================================================================
\section{Conclusions and Outlook}
\label{sec:conclusions}
%==========================================================================

%  !TEX root =  draft.tex
In this work we have analyzed the sensitivity of diffractive gravitational
lensing of Fast Radio Burst (FRB) signals to compact dark matter
objects (MACHOs). 
Diffractive lensing (sometimes called femtolensing or nanolensing in the
literature) means the interference of the two lensed images of a source,
which leads to periodic modulation of the observed frequency spectrum.

We have studied the interplay between the lensing signal and distortions
caused by interstellar scintillation, and we have concluded that the lensing
signal can often be extracted in spite of the distortion. We have developed
a data analysis procedure to extract the lensing signal and showed its
efficiency on simulated data. Our results indicate that the existing
and future FRB surveys carried out by ASKAP, CHIME and SKA will be
able to strongly constrain the cosmological MACHO abundance.
As an
illustration, we studied the sensitivity of an SKA-like survey to
primordial black hole (PBH) dark matter. We concluded that it will be
able to probe the parameter space of PBHs in the mass
range between \SI{e-4}{M_\odot} and \SI{0.1}{M_\odot} and constrain their
abundance down to about 1\% of the total dark matter content of the Universe.

Compared to microlensing searches, which also probe primordial black
holes in this 
mass range, our method is unique because it tests the distribution of dark matter
at cosmological distances rather than in our local cosmic neighborhood.
This distinction is particularly important when generalizing our bounds to other
compact dark matter candidates such as dark matter mini-halos. The local abundance
of dark matter mini-halos may be reduced
due to tidal disruption in the Milky Way,
suppressing the probability of microlensing events.  Moreover, microlensing
searches may be insensitive to dark matter mini-halos due to their non-negligible
spatial extent, which is larger than the Einstein radius at Galactic distances.
At cosmological scales, on the other hand, the Einstein radius is significantly
larger and may well exceed the mini-halo's size, so that it acts effectively
as a point-like lens.  In this case, our results are directly applicable.
Investigating the sensitivity of FRB lensing
to more extended compact objects will be an interesting direction for
future work. 

Apart from dark matter, the range of lens masses considered in this
paper may also contain objects of astrophysical origin, such as brown
dwarfs. However, it appears unlikely that they can produce a
discernible lensing signal in FRB spectra. The reason is that they are
expected to reside in regions of high baryonic density, namely
galaxies, characterized also by high concentrations of free electrons.
The scattering
of radio waves produced by this ambient plasma will be much stronger
than the deflection of light due to gravitational lensing, so that
the effect of the latter will be completely swamped by
scintillation. In principle, the situation might improve for FRBs with
spectra extending to high frequencies $\gtrsim \SI{5}{GHz}$, where the
scintillation effects become weaker. We leave the study of this
interesting possibility for future.
  
Before concluding, we 
emphasize again that the reach of diffractive lensing of radio signals
depends on our understanding of interstellar and intergalactic
scintillation, and thus 
of turbulence in the interstellar and intergalactic medium. While the
scintillation 
parameters we have assumed in this work represent our current best
understanding of these 
phenomena, future observations should allow us to much better quantify the
impact of scintillation. The rapid 
progress in the study of FRBs and in
radio observations in general
will contribute to reducing this uncertainty.

%==========================================================================
\acknowledgments
%==========================================================================

We are grateful to Enrico Barausse, Lam Hui, Matthew Klimek, Sergei
Popov, Igor Tkachev and Sergey Troitsky
for useful discussions. 
JK's work has been partially supported by the European Research Council (ERC)
under the European Union's Horizon 2020 research and innovation program
(grant agreement No.\ 637506, ``$\nu$Directions''). SS is supported by
the Tomalla Foundation.

%==========================================================================
\bibliography{lensing}
\bibliographystyle{JHEP}
%==========================================================================

\end{document}